\documentclass[a4paper,1p]{elsarticle}
\usepackage{graphicx}
\usepackage{epstopdf}
\usepackage{amsmath}
\usepackage{amssymb}
\usepackage{color}

\journal{Physica A}

\newcommand{\SQNORM}[1] { \frac{{#1}}{\sqrt{\pi}} }

\newcommand{\MEANG}[1]{\left [ #1 \right ] }
\newcommand{\MEANMU}[1]{\overline{#1}}

\renewcommand{\exp}[1] { \mbox{e}^{#1}}
\newcommand{\erf}[1] { \mbox{erf}\left ( #1 \right) }

\newcommand{\EXP}[1] { \, \exp{-\zeta^2 I_{#1}^2} }
\newcommand{\ERF}[1] { \, \erf{ \zeta \, I_{#1}} }

\newcommand{\DIFF}[2] { \frac{d{#1}}{d{#2}} }

\newcommand{\CAP}[1] {\caption[#1]{#1}}

\begin{document}
\begin{frontmatter}
\title{
Diversity of scales makes an advantage: the case of the Minority Game
}
%

\author[ada]{%
Miroslav Pi\v{s}t\v{e}k\corref{cor1}}
\ead{miroslav.pistek@gmail.com}
\author[adb]{%
Franti\v{s}ek Slanina}
\ead{slanina@fzu.cz}
\address[ada]{%
The Institute of Information Theory and Automation, 
 Academy of Sciences of the Czech Republic,
Pod~Vod\'{a}renskou~v\v{e}\v{z}\'{i}~4, CZ-18208~Praha
}
\address[adb]{%
Institute of Physics,
 Academy of Sciences of the Czech Republic,\\
 Na~Slovance~2, CZ-18221~Praha
}

\cortext[cor1]{Corresponding author}
\begin{abstract}
We use the Minority Game as a testing frame for the problem of the
emergence of diversity in socio-economic systems. For the MG with
heterogeneous impacts, we show that the direct generalization of the
usual agents' profit does not fit some real-world situations. As 
a typical example we use the traffic formulation of the MG. Taking 
into account vehicles of various lengths it can easily happen that one of the roads
is crowded by a few long trucks and the other contains more drivers
but still is less covered by vehicles. Most drivers are in the shorter
queue, so the majority win. To describe such situations, we
generalized the formula for agents' profit by explicitly introducing 
utility function depending on an agent's impact. Then, the overall profit of the system
may become positive depending on the actual choice of the utility
function.  We investigated several choices of the utility function and showed that this variant 
of the MG may turn into a positive sum game.
\end{abstract}


\end{frontmatter} 

\section{Introduction}

The Minority Game (MG), introduced slightly more than a decade ago
\cite{arthur_94,cha_zha_97,cha_zha_98,jo_ja_jo_che_kwo_hui_98,sa_ma_ri_99}, gained 
the reputation of an ``Ising model'' in the field of Econophysics
\cite{ma_sta_99,bou_pot_03}. It 
is surely an exaggeration, as the complexity of the Ising model is
many orders of magnitude higher, but nevertheless, the MG is a
widely accepted framework for testing a variety of ideas in modelling
socio-economic phenomena. There are several excellent reviews
\cite{joh_jef_hui_03,cha_mar_zha_05,coolen_05,dem_mar_06} of the
vast literature on the subject, which we have no intention to review
again here. The aim of our work is to contribute to understanding the possible 
advantages of diversity, which may be one of the sources of the 
heterogeneity of agents. 

The topic of our work falls within the generalisations of the
canonical MG to the case of several different groups of
agents. Perhaps the most deeply studied is the variant with two types
of agents, called producers and speculators
\cite{sla_zha_99,cha_ma_zha_99,cha_mar_zha_01a}. The market ecology which
emerges, shows fairly clearly that the two groups need each other to
gain better efficiency. In different perspective, the agents may be
divided into groups of leaders and imitators 
\cite{slanina_00,slanina_01a,ang_tor_bas_kor_04,lav_sla_07}. Here
also, the performance of the system as a whole might be increased, if
the imitators are present in proper fraction. If the agents are
allowed to participate in the game or not, depending on their score
\cite{zhang_98,je_ha_hu_jo_00,gia_bou_mez_01,cha_mar_03}, the
bursts of activity reminiscent of the effect of volatility clustering are observed.  
Similar case considers agents who play the game at
certain fixed frequency. Their frequencies may be broadly distributed,
which induces new features into the game
\cite{mar_pia_02,demartino_03,mos_cha_zha_06}. The case of investors with 
different weights was investigated for a very general setup \cite{cha_che_mar_zha_00}. 
In our work, we choose a more specific situation and show that it leads to new non-trivial consequences. 

From the following consideration it can be seen why can be the diversity helpful in
the framework of the MG. In the traffic formulation of the MG, the drivers choose
between two alternative roads and those who
take the less crowded one, are the winners. If all the vehicles have
the same length, only minority of agents can be counted winners,
namely those who drive on the road with minority of drivers. However,
imagine that besides trucks, there are also small city cars. It can
easily happen that one of the roads is crowded by a few long trucks
and the other contains more drivers but still is less covered by
vehicles. Most drivers are in the shorter queue, so the majority 
win.  This is the essence of our paper which is just mathematical 
elaboration of this consideration.

However, the previous simple setting does not take into account 
the utility of the vehicles. In this view, the drivers could choose
infinitesimally short vehicles as optimal which clearly contradicts 
the real traffic situation. Indeed, the longer the car is, the more 
passengers or load it can carry, obviously for the price of higher 
fuel consumption. To model these considerations, we incorporated 
the utility of the vehicles expressed as a function of their 
length. This way we included feature which favour vehicles of finite length.
Nevertheless, the ultimate motivation for the introduction of the utility 
function is the optimization problem the agents face. We consider
the impacts as parameters the agents can choose to maximize their expected utility.

\section{Heterogeneous groups in the MG}

Generally, the MG is a model of the bounded sources
allocation which is driven by agents with bounded
rationality. In the simplest setting, they do not have but two
options. In each round, they aim to choose the right action, 
i.e.~the action chosen by the minority of agents. Then, 
only the minority of agents can win for principal reasons. To adapt to 
their environment, each agent has two strategies. They predict the 
next optimal choice considering only a few previous results. Still, there is 
not any space for an adaptation of agents as both their strategies are fixed. 
Therefore, each agent counts the score of each strategy and then follow the 
currently better strategy. 

In the urban traffic, drivers have cars of various lengths. Following this paradigm, we assign each agent an individual impact on the game. It can be directly interpreted as the length 
of the vehicle the agent is driving, however, some more ambitious interpretations are left to the conclusion. Next, we divide the agents into several groups
according the assigned impacts. Formally, we have $G \in \mathbb{N}^+$ groups of agents. The  group $g \in \{ 1,\ldots,G\}$
contains $N_g$ agents and each of them has impact $I_g \in
\mathbb{R}^+$. The total number
of agents is $N=\sum_{g=1}^G N_g$ and the relative size of group $ g $
will be denoted $\lambda_g \equiv N_g/N$. For further convenience, we introduce the following symbol for the
weighted average of an arbitrary function  $ v_g \in \mathbb{R} $ over all groups 
\begin{equation}
 \MEANG{v_g} \equiv
\sum_{g=1}^G{ \lambda_g v_g } \;.
\end{equation} Throughout this paper, we shall refer to an actual values of the groups' ratios $\{\lambda_g\}_g$ and impacts $\{I_g\}_g$ as the configuration of the game.  To restore the canonical MG, it suffices to set $G$, $\lambda_1$ and $I_1$ equal to $1 $. 

As usual in the MG, all agents are subject to the same external
information and all of them are able to distinguish $P$ different
information (or memory) patterns $\mu \in \{ 1, \ldots, P \} $. These patterns
arrive randomly with uniform probability. We stress that throughout this paper
we shall work within batch MG with uniformly distributed random memories,
i.e.~all P patterns occur with equal frequency, the occurrence of patterns is
uncorrelated in time and strategies' scores are updated only after all $P$ patterns
are used. Within such setting we can relatively easily obtain analytical results,
with no need to resort to computer simulations. In so doing, we rely on the results
of Refs. \cite{cavagna} and \cite{cha_mar_04}, which show that MG with random
uniformly chosen memories is very close, although not identical, to the canonical MG.
All important qualitative aspects are the same and the small quantitative difference
diminishes when we approach the critical point. 

In sequel, we shall denote the average of a function $ f^\mu \in \mathbb{R} $ 
taken over all the information patterns $\mu$ as 
$\MEANMU{ f^\mu } \equiv \frac{1}{P} \sum_{\mu=1}^{P} f^\mu$. 
Next, the action prescribed by
strategy $s = \pm 1 $ of $i$-th agent from group $g$, provided the 
information pattern is $\mu$, is denoted by $a^{\mu,s}_{g,i} \in \{ -1, +1 \}$. 
Then, the currently executed action of an agent reads
\begin{equation} 
\label{INDIVIDUALACTION} 
a_{g,i}^{\mu}(t) \equiv a_{g,i}^{\mu,s_{g,i}(t)} 
\end{equation}
where the random variable $ s_{g,i}(t) $  is an agent's current choice of strategy. It will be specified in detail later on. The following combinations
\begin{equation}
\begin{split}  
&\omega_{g,i}^{\mu} \equiv \frac{a^{\mu,+1}_{g,i}
    + a^{\mu,-1}_{g,i}}{2} \\ 
&\xi_{g,i}^{\mu} \equiv \frac{
    a^{\mu,+1}_{g,i} - a^{\mu,-1}_{g,i} 
 }{2} \end{split} 
\end{equation} 
allow us to write the aggregated action of agents from group $g$ as
\begin{equation} 
\label{ATTENDANCEG} 
A^\mu_g(t) \equiv \sum_{i=1}^{N_g} a_{g,i}^{\mu}(t)
= \sum_{i=1}^{N_g} \omega_{g,i}^\mu + s_{g,i}(t) \, \xi_{g,i}^{\mu} \; .
\end{equation}
Next, for the overall impact of all individual actions 
of all agents we have
\begin{equation} 
\label{OVERALLIMPACT}
  B^\mu(t) \equiv \sum_{g=1}^{G} I_g A^\mu_g(t) \; .
\end{equation}
We follow the batch variant of the MG where the difference of strategies' scores $ q_{g,i}(t) $ 
evolves as
 \begin{equation} 
\label{QUPDATE} q_{g,i}(t+1) -  q_{g,i}(t) =
  - \MEANMU{ \xi^{\mu}_{g,i} \, B^\mu(t) } 
\end{equation}
Note that in this definition the update of the strategies' scores of an agent contains
just the variable $\xi^{\mu}_{g,i} = \pm 1$ indicating the chosen side. The 
alternative updating of the score would be proportional to the impact $I_g$ of the agent 
in question. We shall discuss the differences between these two choices later. 

Now, the current strategy choice $ s_{g,i}(t) $ of an agent is governed by the stochastic rule 
\begin{equation}
\label{STRATEGYCHOICE} \mbox{Prob} \left
  \{ s_{g,i}(t) = \pm 1 \right \} =  \frac{1}{2} \, \frac{ \exp{ \pm
      \Gamma \, q_{g,i}(t)}}{\mbox{cosh}\left ( \Gamma 
    \, q_{g,i}(t) \right )} 
\end{equation} 
where the parameter $ \Gamma \in \mathbb{R}^+ $ is usually named as the ``learning rate''. Indeed, the 
higher the value of $ \Gamma $  is, the more an agent trust his current knowledge represented by the value of $ q_{g,i}(t) $. Henceforth, we denote the mean value of any random variable $x =x(s_{g,i}(t))$ with respect to the  probability distribution (\ref{STRATEGYCHOICE}) by angular brackets,  $\left \langle x \right
\rangle$. Especially, for the mean
value of $s_{g,i}(t)$ itself, we
have
\begin{equation} \label{MAGNETISATION} m_{g,i}(t) \equiv \left
  \langle s_{g,i}(t) \right \rangle = \mbox{tanh}( \Gamma \,
  q_{g,i}(t) ) 
\end{equation}
In what follows, we shall refer this value as magnetization in analogy to purely physical systems. 

In the next section, these magnetizations are necessary to evaluate the performances of the agents in the MG. Optimally, we would calculate all $ m_{g,i}(t) $ as functions of time and made our conclusions. To ease this program a bit, we utilize here the pioneering way the MG was firstly solved using the replica method \cite{cha_ma_ze_00,ma_cha_ze_99}. Thus, we search for the value of $ m_{g,i}(t) $ only at the stationary point of the game's dynamics. Such stationary value $ m_{g,i} $ is defined in this manner 
\begin{equation}
\label{STATMAGNETISATION}
m_{g,i} \equiv \mathop{lim}_{T \rightarrow \infty } \frac{1}{T} \sum_{t=1}^T m_{g,i}(t)
\end{equation}
The same convention will be used also for other variables to denote their stationary value.

To find the stationary state, we observe that, for $\Gamma\to 0 $, the evolution of the strategies' scores (\ref{QUPDATE}) implies the following dynamics of magnetizations (\ref{MAGNETISATION}) 
\begin{equation} 
\label{LYAPUNOV}
  m_{g,i}(t+1)-m_{g,i}(t) \simeq - \frac{1}{2\,N} \frac{\Gamma}{I_g}
  \left (1-m_{g,i}^2(t) \right) \, \frac{\partial H}{\partial
    m_{g,i}} 
\end{equation}
with the Lyapunov function
\begin{equation} 
\label{HAMILTONIAN}
\begin{split} 
H(\{m_{g,i}\}_{g,i}, \{a^{\mu,\pm 1}_{g,i}\}_{g,i}) =\,& \MEANMU{ \left( \sum_{g=1}^{G} I_g \langle
  A_{g}^{\mu} \rangle \right)^2} =\\ 
=\,&\frac{1}{4} \overline{ \left( \sum_{g=1}^{G} I_g  \left
(\sum_{i=1}^{N_g} a^{\mu,+}_{g,i} (1 + m_{g,i} ) + a^{\mu,-}_{g,i} (
1 - m_{g,i}  ) \right )\right )^2} 
\end{split}
\end{equation}
which plays the role of the Hamiltonian in subsequent calculations. Next, by calculating the arguments of the minimum of $ H $ we obtain the stationary values of $ m_{g,i} $.

If we make the score updates proportional to the actual impact of an agent, which is the alternative already discussed nearby Eq. (\ref{QUPDATE}), the equation 
(\ref{LYAPUNOV}) would require a modification. However, a short calculation 
reveals that the modification consists only in missing $I_g$ in the denominator on 
the right-hand side of Eq. (\ref{LYAPUNOV}). Most importantly, both definitions result 
in the same Lyapunov function and further course of the calculations 
is therefore identical. Thus, it does not matter whether the prescription 
for the dynamics of strategies' scores (\ref{QUPDATE}) is proportional to an actual 
impact or not. However, what does matter, is the definition of 
the outcome of the game for an agent, as defined later in Eq. (\ref{RHOGI}). The choice (\ref{QUPDATE}) we did 
earlier was taken to make this definition conform with the definition (\ref{RHOGI}) 
(to be seen later).

In fact, the previous approach with Hamiltonian $H$ playing the role of Lyapunov function is not the most general. To determine the behaviour of the MG for arbitrary $ \Gamma $, there exists another
 path of calculations mastering stochastic calculus \cite{mar_cha_01a}. However,
 it was shown that finite value of $ \Gamma $ influences the results only in the 
non-ergodic phase while in the ergodic phase the results are independent
 of $ \Gamma $. Here we investigate only the ergodic phase, thus the limit $ \Gamma \rightarrow 0 $ is acceptable and the use of Lyapunov function is justified. 

In the course of computations, we shall work in the thermodynamic 
limit of infinite number of agents, $N\to\infty$, keeping all group ratios $ \lambda_g $ fixed,
and at the same time limiting the number of information patterns to infinity, $P\to\infty$, holding 
the fraction   
\begin{equation} 
\label{ALPHA} 
\alpha \equiv \frac{P}{N}
\end{equation}
fixed. This value is the principal parameter of both canonical and generalized MG. 

\section{The profit of agents}

In this section, we have to detour from the conventional notation used in 
the context of the MG. To incorporate the utility function of agents, it is better to write about their profits instead of their losses. To establish the connection to the canonical MG, a clear relation to volatility 
\begin{equation} \label{SIGMA} \sigma^2 \equiv 
  \frac{1}{N} \mathop{lim}_{T \rightarrow \infty } \frac{1}{T} \sum_{t=1}^T  \MEANMU{ \left\langle B^\mu(t)^2 \right\rangle }  \;,
\end{equation}
which is the usual characteristic of agents' performance, will be derived.

First of all, we define the current outcome of an individual agent $i$ of group $g$ as \begin{equation} 
\rho_{g,i}(t) \equiv  -
\MEANMU{ \left\langle a_{g,i}^\mu(t) B^\mu(t) \right\rangle } \; .
\label{RHOGI} 
\end{equation}  
It indicates whether an agent was successful in the current round of the batch MG. 
His ability to obey the ``join-minority imperative'', i.e. to choose $ a_{g,i}^\mu(t) = -\mbox{sign}(B^\mu(t)) $ for a typical information pattern $ \mu $, guarantees him positive outcome $ \rho_{g,i}(t) $ and the same holds vice versa. We note that the fact that an agent's impact $I_g$ is not included in formula (\ref{RHOGI}) is the crucial point of our modification of the MG. 

Next, for the average outcome within the group $g$ we have
\begin{equation}
\label{RHOG}
\rho_g(t) \equiv \frac{1}{N_g} \sum_{i=1}^{N_g} \rho_{g,i}(t)  = - \frac{1}{N_g} \,	  \MEANMU{ \left\langle A^\mu_g(t) B^\mu(t) \right\rangle } \;.
\end{equation} 
Coming to the stationary state, we can write (the stationary value of) the overall average outcome $ \rho $ as
\begin{equation}
\label{RHO}
\rho \equiv \MEANG{ \rho_g }
\end{equation} 

To relate an average outcome within a group to an average profit of that group, we have to take into account the utility function $U(I)$. We follow the standard way it is defined in classical economics.
Its purpose is to quantify agents' preferences over the whole spectra of possible impacts \cite{bar_hamm_seid}. In other words, the more an agent prefers impact $ I $, the higher value is assigned to that impact by $ U(I) $. Such utility function can be interpreted, for instance, as the profit of an owner of the vehicle of length $ I $. For the sake of simplicity, we assume that the utility function is common for all agents of all groups and non-negative. Now, we define the average profit $ \Pi_g^{\{U(I)\}} $ of group $ g $ provided the utility function is $ U(I) $ in this manner \begin{equation} 
\Pi_g^{\{U(I)\}}  \equiv U(I_g)\,
    \rho_g \;.
\end{equation}
Then, the overall average profit over all groups reads 
\begin{equation} 
\label{PROFIT}
\Pi^{\{U(I)\}}  \equiv \MEANG{ \Pi_g^{\{U(I)\}} }  \;.
\end{equation}
We note that introduction of $ U(I) $ is by no means equivalent to readjusting of the impacts. 

If we consider all impacts to have the same utility, i.e. $ U(I) = 1 $, we obtain an extremal case of profit $ \Pi^{\{1\}} = \rho $. On the other hand, prescribing an utility of impact directly proportional to its size, i.e. $ U(I) = I $, we observe $\Pi^{\{I\}} = -\sigma^2$, so the minus volatility (\ref{SIGMA}) is another extremal case of the profit defined by formula (\ref{PROFIT}). Realistic situations will lie
somewhere in the middle. For example, we can choose $U(I)=I^d$ for
$0<d<1$ or $ U(I) = I / (I+1) $.  However, in the following we shall
mainly compare the extremal cases $- \sigma^2$ and $\rho$, to better 
see the difference. We shall see
that this shift of focus has important implications on the conclusions
we shall draw from the results of the computations.

At the moment, we define the mutual predictability between groups $f,g \in \{ 1,\ldots,G\}$ 
\begin{equation} 
\label{THETAFG} \theta_{fg} \equiv
  \frac{N}{N_f N_g} \, \MEANMU{ \langle A_f^\mu \rangle \langle
    A_g^\mu \rangle } 
\end{equation}  
which is a direct generalisation of the usual definition of the predictability in the canonical MG \cite{cha_mar_zha_05}. It corresponds to the ability of group $ f $ to adapt to the aggregated attendances $ A_g $, and vice versa. The less the value of $ \theta_{fg} $ is, the more the groups $ f,g $ are adopted to each other as their aggregated attendances $ A_f^\mu $ and $ A_f^\mu $ are more anti-correlated. Then, for the mean predictability of group $ g $ it holds 
\begin{equation} \label{THETAG} \theta_g \equiv \frac{1}{N} \, \MEANMU{ \langle A_g^\mu \rangle \langle  B^\mu \rangle }  =
\MEANG{ I_f \theta_{fg}} \; .
\end{equation} 
Its knowledge is the key ingredient in the calculation 
of $\rho_g$. Indeed, if we apply the approximation 
\begin{equation} 
\sum_{f=1}^G I_f \sum_{i=1}^{N_g}
  \sum_{j=1}^{N_f} \MEANMU{ \xi_{g,i}^{\mu} \, \xi_{f,j}^{\mu}  }
  \langle ( s_{g,i} - m_{g,i} ) (s_{f,j} - m_{f,j} ) \rangle  \simeq
  I_g \sum_{i=1}^{N_g} \MEANMU{ ( \xi_{g,i}^{\mu})^2 } ( 1 - m^2_{g,i}
  )  
\end{equation}
which is justified in the ergodic phase, see again Ref. \cite{cha_mar_zha_05},
introduce the set of order parameters for $ g \in \{1, \ldots, G \} $
\begin{equation} Q_g \equiv \frac{I_g^2}{N_g}
   \sum_{i=1}^{N_g} m^2_{g,i} \; ,
\label{QG} 
\end{equation}
and compare definitions of $ \rho_g $ and $ \theta_{g} $, see (\ref{RHOG}) 
and (\ref{THETAG}), respectively, we obtain
 \begin{equation} 
\label{RHOG2THETAFG} 
\rho_g  = \frac{Q_g - I_g^2}{2 I_g} - \theta_{g} \;.
\end{equation} 
At this moment, we can realise the importance of $\theta_{g}$. It is not only 
a characteristic of groups' ability to predict the overall impact $ B^\mu $, 
 but it is also a link connecting the knowledge of the outcome $\rho_g$ of group $ g $ to the knowledge of the minima of Hamiltonian $H$. Recalling the definition of $ \theta_{fg} $, see (\ref{THETAFG}), 
together with equation (\ref{HAMILTONIAN}), we can derive that 
\begin{equation} \label{THETAFG2H} \theta_{fg} = \frac{1}{2} \, \frac{N}{N_f N_g} \,
  \frac{d^2 H}{dI_f \, dI_g} \;.\end{equation} 
Thus, as soon as we know $H$ as function of the impacts $ I_g $, $ g \in \{ 1, \ldots, G \} $, we deduce also all mean predictabilities $ \theta_g $ for all groups using the mutual predictabilities $ \theta_{fg} $.

\section{Replica solution}

Whatever utility function $ U(I) $ we choose for the impact of the agents, the
key elements for the calculation are the average groups' outcomes $\rho_{g}$ which are
functions of the stationary magnetizations $m_{g,i}$, see (\ref{RHOG2THETAFG}). To obtain
these magnetizations, it suffices to minimise the Hamiltonian $H(\{m_{g,i}\}_{g,i},  \{a^{\mu,\pm 1}_{\mu,g,i}\}_{g,i} )$ as it
is also Lyapunov function for their dynamics, remember equation (\ref{LYAPUNOV}). 
The technical obstacle to actual computation consists in the quenched disorder contained 
in the Hamiltonian
(\ref{HAMILTONIAN}) due to the random strategies $a_{g,i}^{\mu,\pm 1}$. The standard procedure to tackle this 
complication is the replica method  \cite{me_pa_vi_87,dotsenko_95}.

To briefly depict this method, consider a system with Hamiltonian $H_a(m)$ dependent on state variable $m$ and also on random, but quenched disorder $a$ distributed with probability $P(a)$. In that setting, even the partition function $Z_a = \int \mathrm{e}^{-\beta\,H_a(m)} \, \mathrm{d}m$ is a random variable and so is the free energy $F_a = - \frac{1}{\beta} \, \mbox{log} \, Z_a$.  Then, the physically relevant quantity is the mean free energy $F \equiv \int F_a \, P(a) \, \mathrm{d}a$. To ease its computation, the replicated partition function $Z(n)$ is introduced as the central notion of this method. It is calculated by replicating the system $n$-times with the same disorder $a$, and finally, by taking an average over this disorder 
\begin{equation}
 Z(n) \equiv  \int Z_a^n P(a) \, \mathrm{d}a = \int \mathrm{e}^{-\beta \, n \,F_a}  P(a) \, \mathrm{d}a
\end{equation}
As the second equality indicates, $Z(n)$ is also a characteristic function for the random variable $F_a$.
Thus, for the mean free energy it holds
\begin{equation}
\label{REPLICATEDPARTITION}
 F  = - \frac{1}{\beta}
\mathop{lim}_{n \rightarrow 0} \DIFF{Z(n)}{n}  
\end{equation}
However, this limiting process is only formal as we are unable to calculate $Z(n)$ for general $n$, except for positive integers. The existence and uniqueness of an analytical continuation of $Z(n)$ is an open question of the contemporary mathematical physics. 

Full solution of the canonical MG using
replica method is available \cite{cha_ma_ze_00,ma_cha_ze_99}, as well
as solution of various modifications of the MG
\cite{cha_ma_zha_99,cha_mar_zha_01a,mar_pia_02,mos_cha_zha_06,dem_mar_00,dem_gia_mos_03,cha_dem_mar_per_06}.
 We apply this method straightforwardly to our case. For the replicated partition function we obtain formula
\begin{equation}
\label{REPLICATEDPARTITIONOFOURSYSTEM} Z(n) =
\sum_{ \big \{ a_{g,j}^{\mu,s} \big \} \in \big \{ -1,+1 \big \}^{2PN} } \left( \int_{-1}^1 \mathrm{e}^{-\beta\,H(\{m_{f,i}\}_{f,i},\{a^{\mu,\pm 1}_{g,j}\}_{g,j})} \prod_{f,i}\mathrm{d}m_{f,i} \right )^n 
\end{equation}  
where the sum goes over all possible realizations of the quenched disorder. Using Hubbard-Stratonovich transformation, we introduce auxiliary variables $z^a \in \mathbb{R} $ indexed by replica index $a \in \{ 1, \ldots, n \}$. Thus, we obtain the previous formula in the following form 
\begin{equation}Z(n) = \int_{-1}^1 \! \left (
\int_\mathcal{R} \! \exp{
-\frac{1}{2} \sum_{a=1}^n (z^a)^2 - \frac{\beta}{2P}
\sum_{a,b=1}^n z^a z^b \sum_{g,i} {  I_g^2 \left ( 1 + m^a_{g,i} m^b_{g,i}  \right )}}  \prod_a \mathrm{d}z^a   \right )^{\!P} \!\!\!\! \prod_{a,g,i} \! \mathrm{d}m^a_{g,i}
\end{equation} 
To compute it, it is necessary to introduce the replica-symmetric ansatz and then 
apply the saddle-point method. Next, we calculate the mean free energy $F$ using equation (\ref{REPLICATEDPARTITION}). Finally, to achieve the formula for the minima 
of Hamiltonian $H$, it 
suffices to carry out the following limit 
\begin{equation} 
H = \mathop{lim}_{\beta \rightarrow \infty} F
\end{equation} 

Without resorting more details of the computation, we note that at
the core there is an equation for an auxiliary quantity $\zeta \in \mathbb{R^+}$ parameterized by $ \alpha $ defined in (\ref{ALPHA})
\begin{equation} \label{FINALALPHA} \MEANG{ 4 \zeta^2 I_g^2  - 2 \zeta^2 I_g^2 \ERF{g} + \ERF{g} - \SQNORM{2 \zeta I_g}  \EXP{g}} = \alpha \;.
\end{equation}
Once $\zeta$ is found, we can calculate the
order parameters
\begin{equation} \label{FINALQG} Q_g = I_g^2  - I_g^2 \ERF{g} +
  \frac{1}{2 \, \zeta^2} \ERF{g} - \SQNORM{1} \frac{I_g}{\zeta}
  \EXP{g}  \; ,
\end{equation}  
 the minimum value of the Hamiltonian
\begin{equation}
 \label{FINALH} H = \frac{N}{2}  \MEANG{Q_g +
    I_g^2} \,\left ( 1 - \frac{\MEANG{ \ERF{g} }}{ \alpha} \right
  )^2
\end{equation} 
and hence the mutual predictability using relation (\ref{THETAFG2H}) reads
\begin{equation} 
\begin{split}
 \label{FINALTHETAFG} 
\theta_{fg} = \,&\frac{1}{\alpha} \left ( 2 \,
\zeta I_g - \zeta I_g \ERF{g} - \SQNORM{1}\EXP{g} \right ) \times \\
&\times\left ( 2 \,
\zeta I_f - \zeta I_f \ERF{f} - \SQNORM{1}\EXP{f} \right )  +  \\ &
\frac{\alpha - \MEANG{\ERF{h}}}{2 \alpha } \left (   \frac{2 - 
  \ERF{f} }{\lambda_f} \, \delta_{fg} - I_g I_f \frac{(2 - \ERF{g})(2
  - \ERF{f})}{\MEANG{ 2 I_h^2 - I_h^2 \ERF{h}} } \right ) \;.
\end{split}
\end{equation}
 Considering equation (\ref{RHOG2THETAFG}), we
conclude 
\begin{equation} 
\label{FINALRHO}
\rho_{g} = \frac{\ERF{g}}{4 \, \zeta^2 \, I_g} + \frac{\MEANG{\ERF{f}} 
}{2 \, \alpha \, \zeta } \left ( 2 \zeta I_g - \zeta I_g \ERF{g} -
\SQNORM{1} \EXP{g} \right ) - I_g\;. 
\end{equation} 
The expression (\ref{FINALRHO}) is the main result we shall 
build on further in the following sections. Before going to that, we
complete the calculations by the formula for the volatility
(\ref{SIGMA}) which is
\begin{eqnarray} 
\label{FINALSIGMA} \sigma^2 =
  \MEANG{I_g^2} + \frac{{\MEANG{ \ERF{g}}}^2}{4 \alpha \,\zeta^2}  -
  \frac{\MEANG{ \ERF{g}}}{2 \,\zeta^2}\;. 
\end{eqnarray} In the following two sections, we shall analyse these
results first at the critical point and then deep in the ergodic
regime. 

\section{Critical behaviour}

For the minimization of $ H $ the analysis of critical behaviour is crucial. 
The calculation of (\ref{REPLICATEDPARTITIONOFOURSYSTEM}) is valid only 
in the ergodic phase of the MG as it was already examined in Ref. \cite{cha_mar_zha_05}. Therein, 
the formula for susceptibility $ \chi $ is
obtained which in our setting directly generalizes to \begin{equation} 
\chi = \frac{ \MEANG{ \ERF{g}} }{ \alpha - \MEANG{\ERF{g}}}
\end{equation} And it is the very fact of diverging susceptibility
what characterises the critical point. Thus, it was deduced that 
the ergodic phase of the MG corresponds to values $\alpha 
\ge \alpha_{c}$ with the critical value $ \alpha_c $ obtained from 
\begin{equation} \label{ALPHAC} \alpha_{c} = \MEANG{\, \erf{
      \zeta_{c} \, I_g }}
\end{equation} 
with $\zeta_{c}$ given by an
equation emerging from the previous equation together with 
(\ref{FINALALPHA}) 
\begin{equation}
\label{ZETAC}\MEANG{ 2 \, \zeta_{c}^2 \,
I_g^2  - \zeta_{c}^2 \, I_g^2 \, \erf{
\zeta_{c} \, I_g } - \SQNORM{1} \, \zeta_{c}
\, I_g \, \, \exp{- \zeta_{c}^2 \, I_g^2 }} = 0 \end{equation} 
Considering only one group of players, critical
coefficient $\alpha_{c} \simeq 0.3374$ of the canonical MG
reappears. Generally, the critical
value $ \alpha_c $ is equal or below this ``canonical'' value for any game configuration $\{\lambda_g,I_g\}_g$ due to
the concavity of $ \erf{x} $ in (\ref{ALPHAC}). 

In the canonical MG, the predictability vanishes at the critical
point $ \alpha = \alpha_c $. This observation does not hold for the
mutual predictabilities $ \theta_{fg} $ directly, however, it holds for the mean 
predictability $ \theta_{g} $ of an arbitrary group $ g $. To verify it, we substitute 
condition (\ref{ALPHAC}) for $ \alpha_c $ into the definition of $ \theta_g $, see (\ref{THETAG}), 
using equation (\ref{FINALTHETAFG}). The achieved equation for $ \theta_g $ at the critical point 
is directly proportional to the left side of (\ref{ZETAC}) and on that account $ \theta_g $ does vanish here.

Another significant characterization of the critical point in the canonical MG was related to the notion of frozen agents firstly mentioned in Ref. \cite{cha_mar_99}. An agent $i$ of group $g$ is called frozen if one of his strategies is significantly better than the other. Thus, in the stationary point it holds $m_{g,i} = \pm 1$. In the canonical MG the fraction of frozen agents $\phi$ is maximized at the critical point and it decreases with increasing $\alpha > \alpha_c$. For the fraction of frozen agents $\phi_g$ within group $g$ in the MG with heterogeneous impacts we obtain 
\begin{equation}
\phi_g = 1 - \erf{ \zeta I_g }
\end{equation} 
following Ref. \cite{bru_zec_94}. Then, the overall fraction of frozen agents is
\begin{equation}
\label{MEANPHI}
\phi = 1 - \MEANG{\erf{ \zeta I_g }}\;.
\end{equation} 
From equation (\ref{FINALALPHA}) we can see that $\zeta$ grows with increasing $\alpha$. Thus, even in the MG with heterogeneous impacts the overall ratio of frozen agents $\phi$ is maximized at $\alpha = \alpha_c$ and decreases with increasing $\alpha > \alpha_c$. Furhtermore, this observation holds regardless of the actual game configuration $\{\lambda_g,I_g\}_g$.

Also, the loss in the canonical MG, which is represented by volatility $ \sigma^2 = - \Pi^{\{I\}}$, was 
minimized at the critical point.  We will show that this fact is violated in the generalized MG. 
Substituting the condition 
(\ref{ALPHAC}) for $ \alpha_c $ into the 
formula for a group outcome $ \rho_g $, remember (\ref{RHOG}), we obtain 
a simplified formula at the critical point
\begin{eqnarray}
\rho_{g} & = & \frac{1}{2} \, \erf{\zeta_{c} I_g} \left ( \frac{1}{2 \, \zeta_{c}^2 \, I_g} \, -  I_g \right ) - \frac{1}{ 2 \, \zeta_{c} \sqrt{\pi} } \, \exp{-\zeta^2_{c} I^2_g}
\end{eqnarray} 
We can show that $\rho_{g} \le 0$ holds for all groups $g$ and all the possible game configurations $\{\lambda_g,I_g\}_g$ as long as we are at the critical point. It
implies that even the average profit $ \Pi^{\{U(I)\}}$ is negative here for an
arbitrary utility function $ U(I)$ (which is non-negative by definition). Nevertheless, in the next
section, we will see that there exist game configurations where $
\Pi^{\{U(I)\}} > 0 $ holds for some specific utility functions away
from the critical point. The existence of positive average profits away 
from critical point implies that the critical point is no more optimal for the 
generalized variant of the MG. However, this observation depends on the considered 
utility function.

\section{Two groups}

Next, we examine the easiest generalisation of the canonical MG. We
consider a game consisting of two groups of agents, $G=2$, and depict
their co-habitation. Firstly, we fix a scale of impacts
$I_g$ \begin{equation} \label{IMPACTSCALE} \MEANG{I_g^2} =
  1. \end{equation} This particular choice establishes a relation
to the canonical MG as $\MEANG{I_g^2}$ corresponds to volatility
$\sigma^2$ in the game of randomly guessing agents. Thus, having this
normalisation, we work on the scale of the canonical MG. We also prefer this normalisation 
because the calculations for the game with infinitely many groups 
are valid only provided $ \MEANG{I_g^2} < \infty $ \cite{cha_che_mar_zha_00}. 

\begin{figure}[h!t] \centering
\includegraphics[scale=0.5,clip]{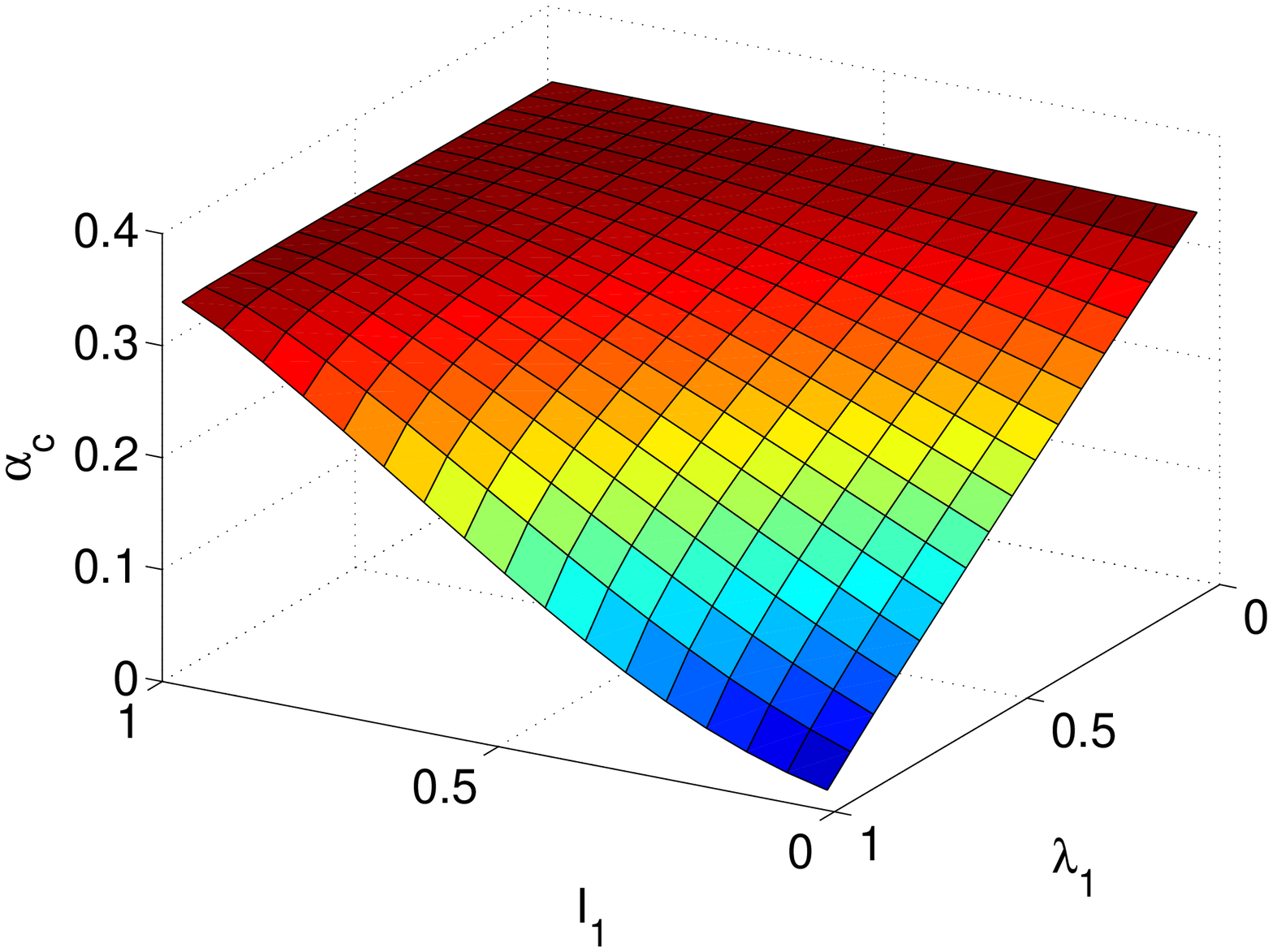}
\CAP{Critical alpha $\alpha_c(\lambda_1,I_1)$ calculated for the game of two groups ($ G=2 $) with groups' impacts normalized as $ \MEANG{I_g^2} =1 $.}
\label{FIG.ALPHAC} 
\vfill
\includegraphics[scale=0.5,clip]{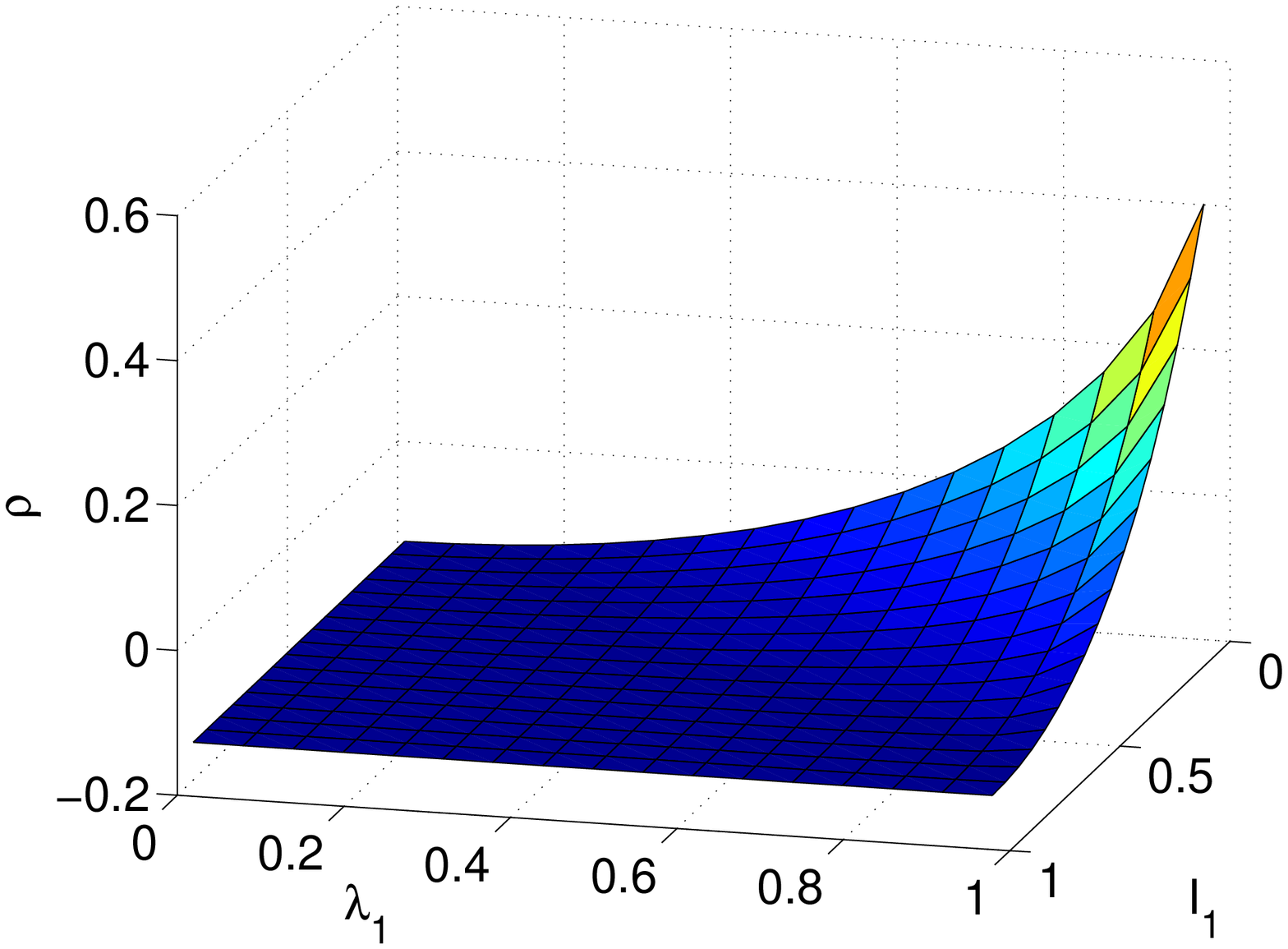}
\CAP{The average overall outcome $\rho(\lambda_1,I_1)$ calculated for the game of two groups ($ G=2 $) with $\alpha = 0.4$ and groups' impacts normalized as $ \MEANG{I_g^2} =1 $.}
\label{FIG.RHO}
\end{figure}

In all the subsequent figures it suffices to consider impact $ I_1 \in [0,1]$ since the rest of 
function values, i.e. those of argument $I_1 > 1$, are determined through (\ref{IMPACTSCALE})
 by reversing the groups. As for the variable $ \lambda_1 $, we restrict ourselves to $\lambda_1 \in [0,1) $. 
It is possible to prove that $ \alpha_c(\lambda_1,I_1) $ is discontinuous 
along the line $ \lambda_1 = 1 $ considering the normalisation (\ref{IMPACTSCALE}). 
In Fig. \ref{FIG.ALPHAC}, it is clearly visible that
the critical value $\alpha_{c}(\lambda_1,I_1)$ is, indeed, lower or equal
to the critical value $\alpha_{c} \simeq 0.3374$ observed in the canonical MG. We picked $\alpha = 0.4$ for all subsequent calculations in order to remain safely in the ergodic phase as it was discussed in the previous section. 

\begin{figure}[h!t] \centering
\includegraphics[scale=0.5,clip]{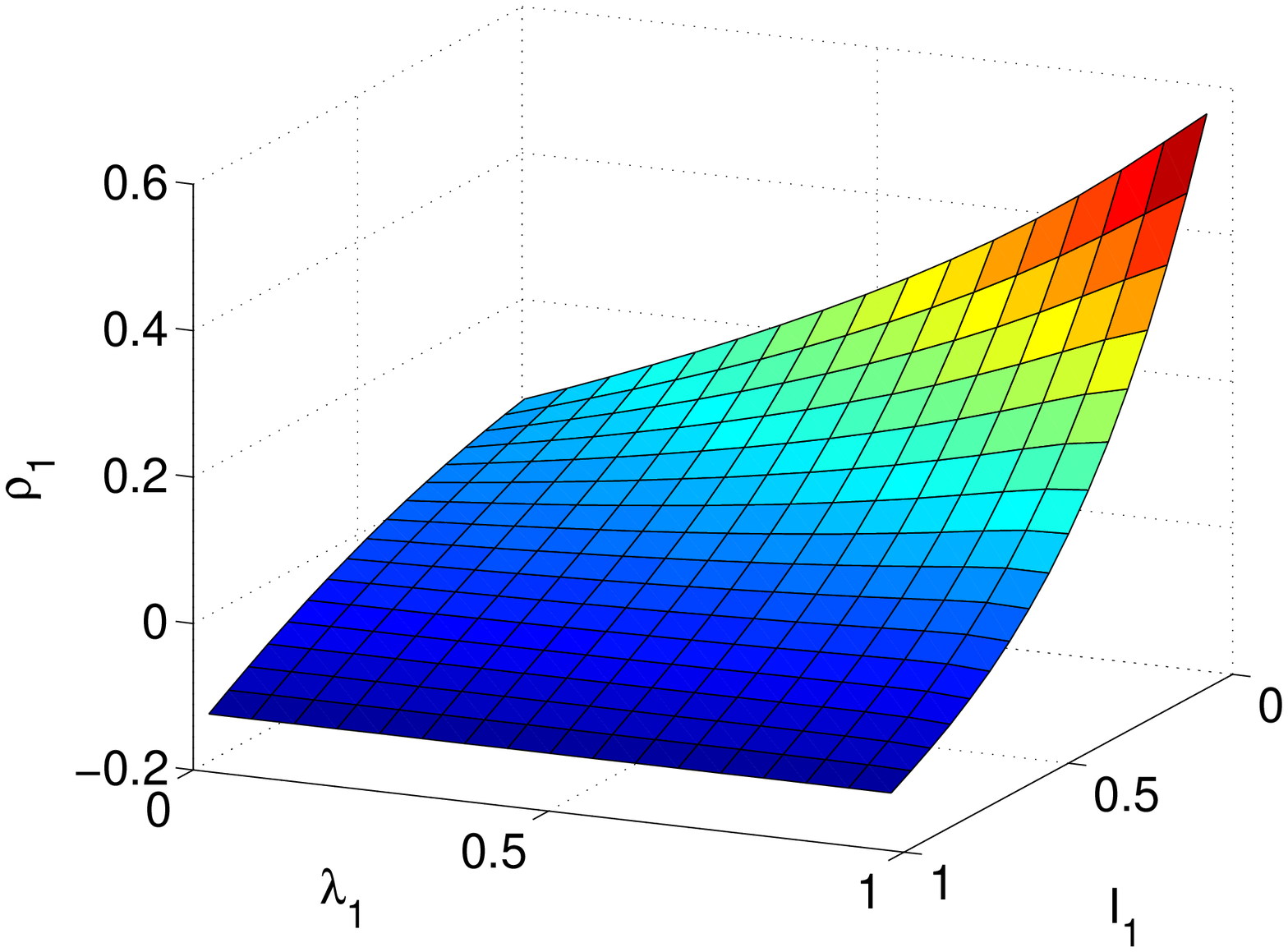}
\CAP{The average outcome $\rho_1(\lambda_1,I_1)$ of the first group calculated for the game of two groups ($ G=2 $) with $\alpha = 0.4$ and groups' impacts normalized as $ \MEANG{I_g^2} =1 $.}
\label{FIG.RHO1} 
\vfill
\includegraphics[scale=0.5,clip]{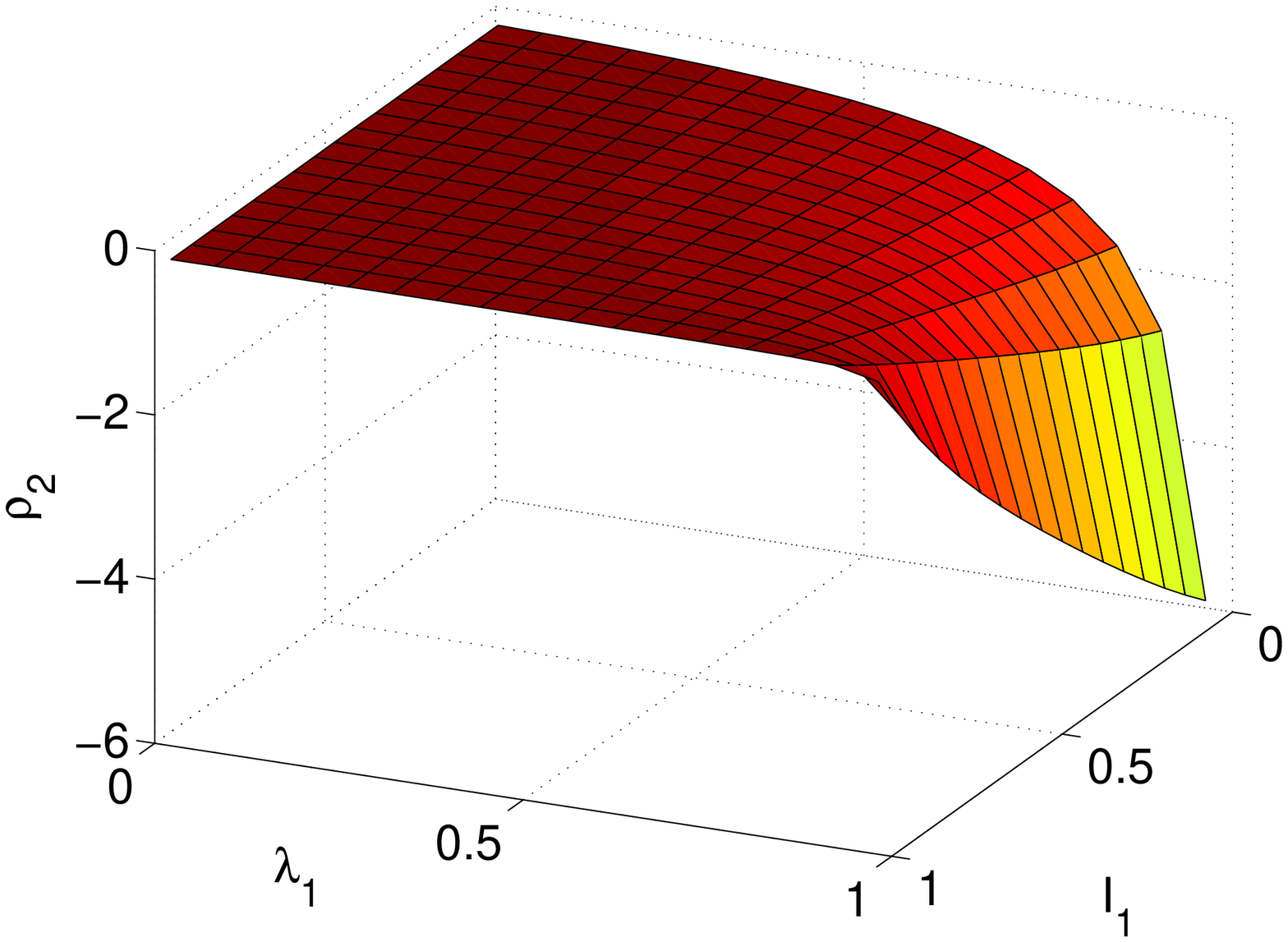}
\CAP{The average outcome $\rho_2(\lambda_1,I_1)$ of the second group calculated for the game of two groups ($ G=2 $) with $\alpha = 0.4$ and groups' impacts normalized as $ \MEANG{I_g^2} =1 $.}
\label{FIG.RHO2}
\end{figure}

The principal novelty in our result is that for some
game configurations $\{\lambda_g,I_g\}_g$ the overall average 
outcome $\rho = \Pi^{\{1\}}$ can even be positive as depicted in Fig. \ref{FIG.RHO}. In other words, the
majority of agents wins in these cases. From Fig. \ref{FIG.RHO1} and
Fig. \ref{FIG.RHO2} showing separate outcomes of each group $\rho_1$ and 
$\rho_2$, respectively, we can deduce what is behind this phenomena. To gain on
average, one group has to sponge off the other. In the case of two
groups, the group with smaller impact is always winning. 
Its positive outcome is not very high, in fact, but when weighted by the 
groups' ratios, it can overweight negative outcome of the other group.

\begin{figure}[h!t] \centering
\includegraphics[scale=0.5,clip]{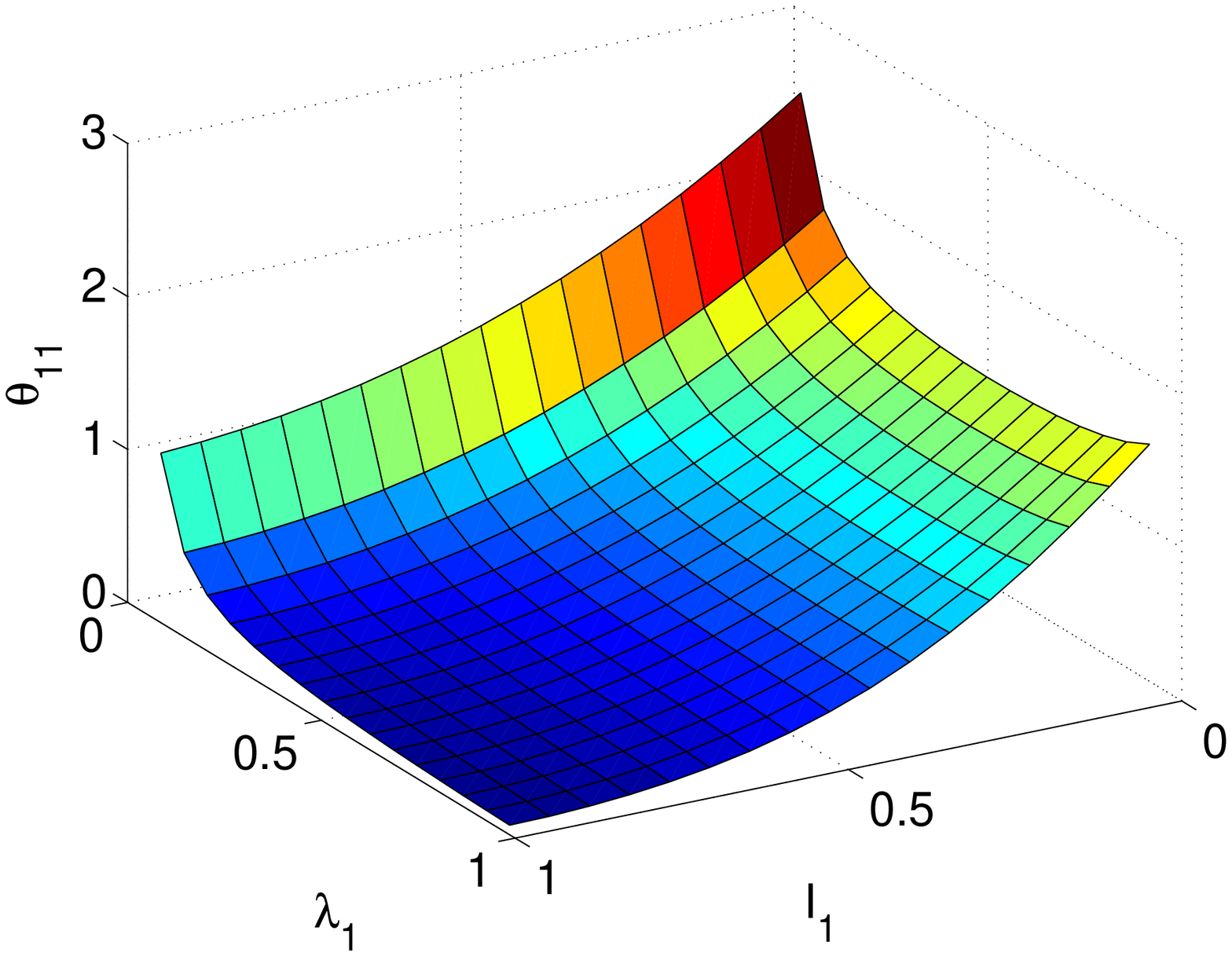}
\CAP{The predictability $\theta_{11}(\lambda_1,I_1)$ of the first group calculated for the game of two groups ($ G=2 $) with $\alpha = 0.4$ and groups' impacts normalized as $ \MEANG{I_g^2} =1 $.}
\label{FIG.THETA11}
\vfill
\includegraphics[scale=0.5,clip]{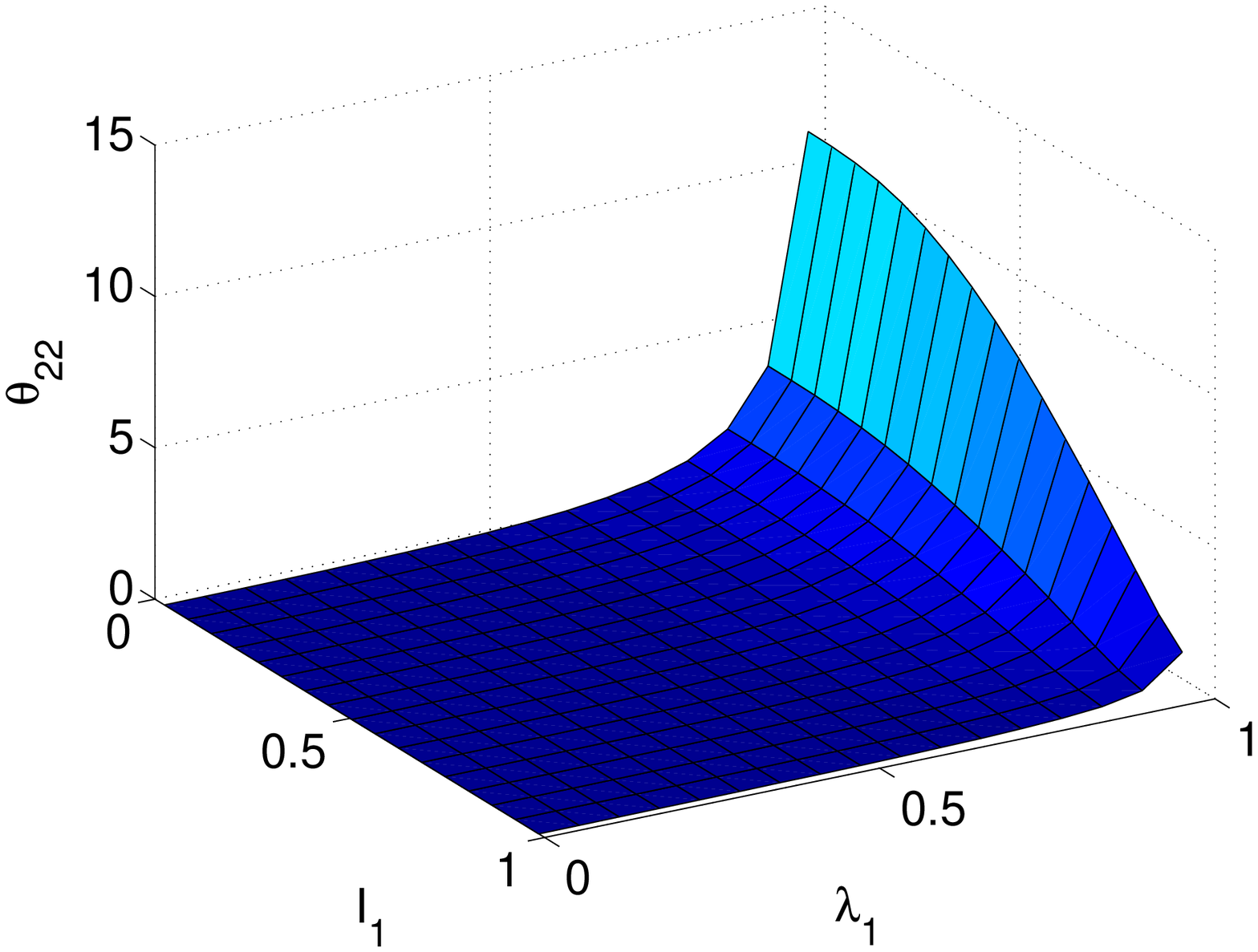}
\CAP{The predictability $\theta_{22}(\lambda_1,I_1)$ of the second group calculated for the game of two groups ($ G=2 $) with $\alpha = 0.4$ and groups' impacts normalized as $ \MEANG{I_g^2} =1 $.}
\label{FIG.THETA22}
\end{figure}

\begin{figure}[h!t] \centering
\includegraphics[scale=0.5,clip]{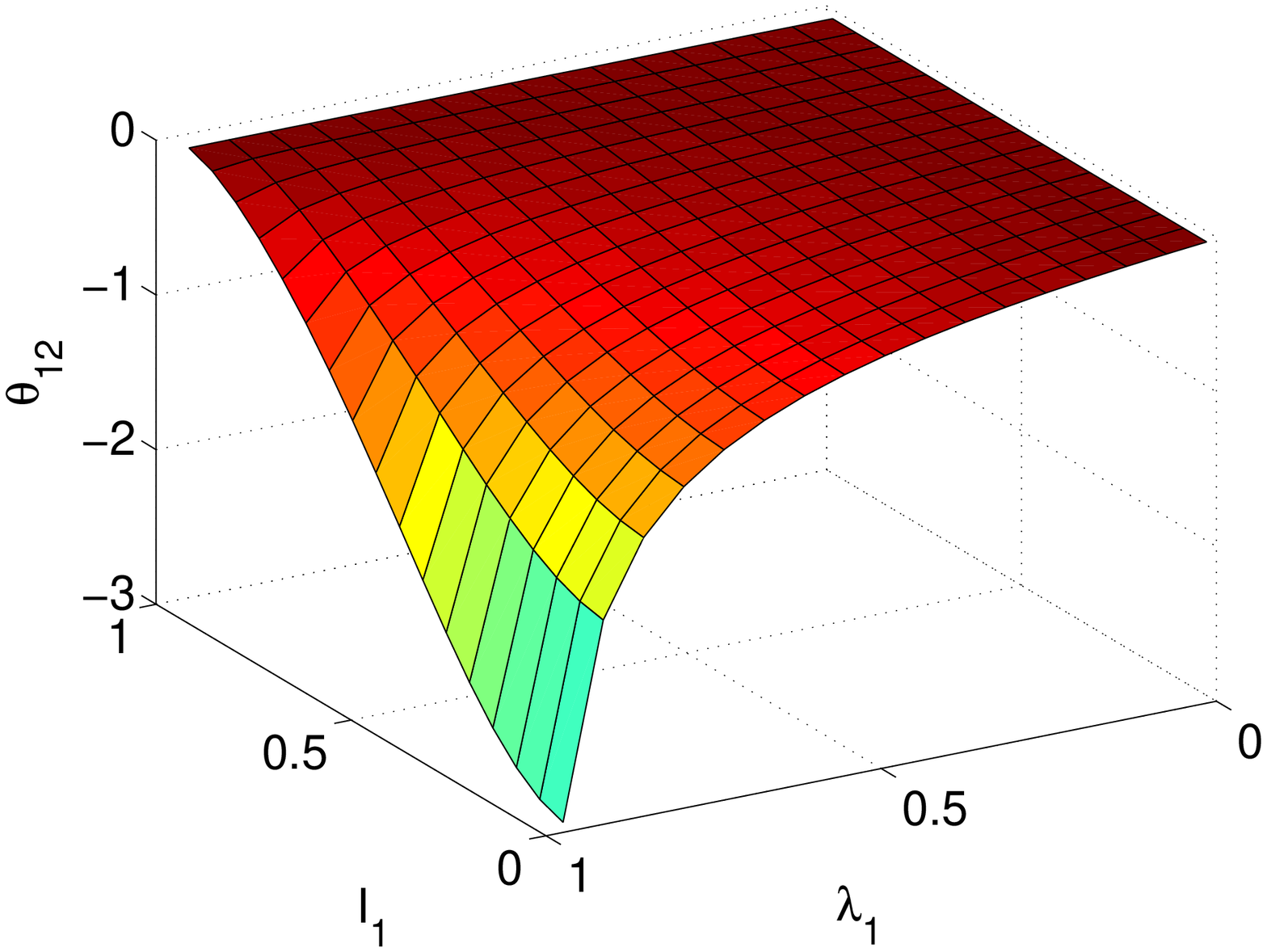}
\CAP{The mutual predictability $\theta_{12}(\lambda_1,I_1)$ between groups calculated for the game of two groups ($ G=2 $) with $\alpha = 0.4$ and groups' impacts normalized as $ \MEANG{I_g^2} =1 $.}
\label{FIG.THETA12}
\vfill
\includegraphics[scale=0.5,clip]{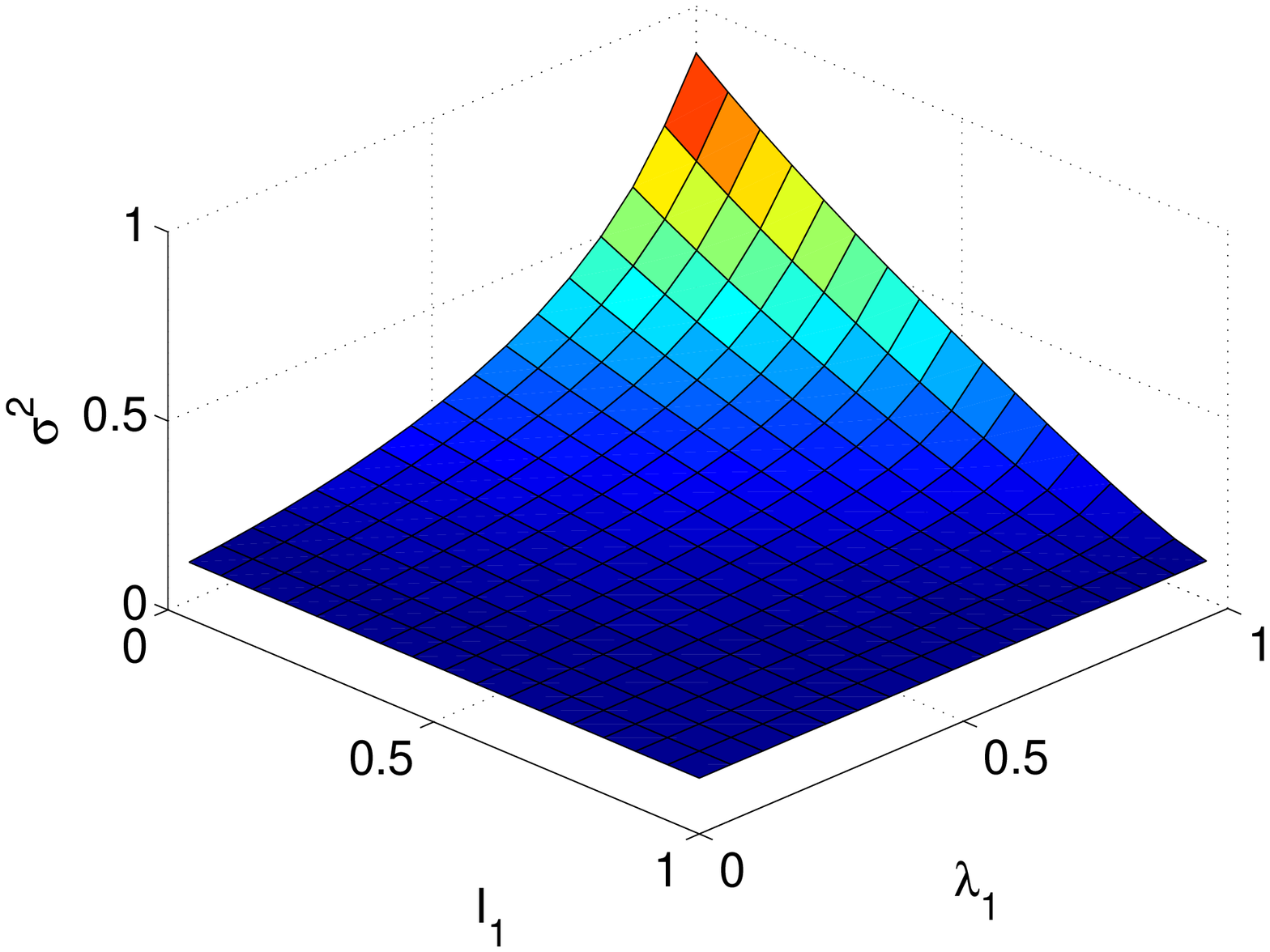}
\CAP{The volatility $\sigma^2(\lambda_1,I_1)$ calculated for the game of two groups ($ G=2 $) with $\alpha = 0.4$ and groups' impacts normalized as $ \MEANG{I_g^2} =1 $.}
\label{FIG.SIGMA} 
\end{figure}

In the next three figures, we plot the predictabilities $\theta_{11}$, 
$\theta_{22}$ and mutual 
predictability $\theta_{12}$, see Fig. \ref{FIG.THETA11}, Fig. \ref{FIG.THETA22}
and Fig. \ref{FIG.THETA12}, respectively. Note that the positiveness of the diagonal elements 
of $ \theta_{fg} $ is enforced by their definition, see (\ref{THETAFG}).
Next, we observe that $ \theta_{12} $ is negative for all configurations of the game. 
In other words, the aggregated actions $ A^\mu_1 $ and $ A^\mu_2 $ are anticorrelated
for the MG consisting of two groups.

In Fig. \ref{FIG.SIGMA}, we show the volatility $\sigma^2 = - \Pi^{\{I\}}$ 
defined by (\ref{SIGMA}). The closer $\sigma^2$ is to
$1$, the closer is the result of the game to the behaviour of
randomly guessing agents. We observe that this holds in the vicinity
of $\lambda_1 = 1$ and $I_1 = 0$. In other words, the co-adaptation of
agents is poor for these game parameters. 
However, it is the same area where $\alpha_{c}$ falls almost
to $0$, confer Fig. \ref{FIG.ALPHAC}. It resembles a similar feature of the canonical
MG, where $ \sigma^2 $ grows with $\alpha$
receding from $\alpha_{c}$. Here, the situation is
analogous. Even though $\alpha$ was fixed in this set of calculations,
$\alpha_{c}$ is decreasing in the region of our interest
and thus the distance $\alpha - \alpha_{c}$ is growing. 

\begin{figure}[h!t] \centering
\includegraphics[scale=0.5,clip]{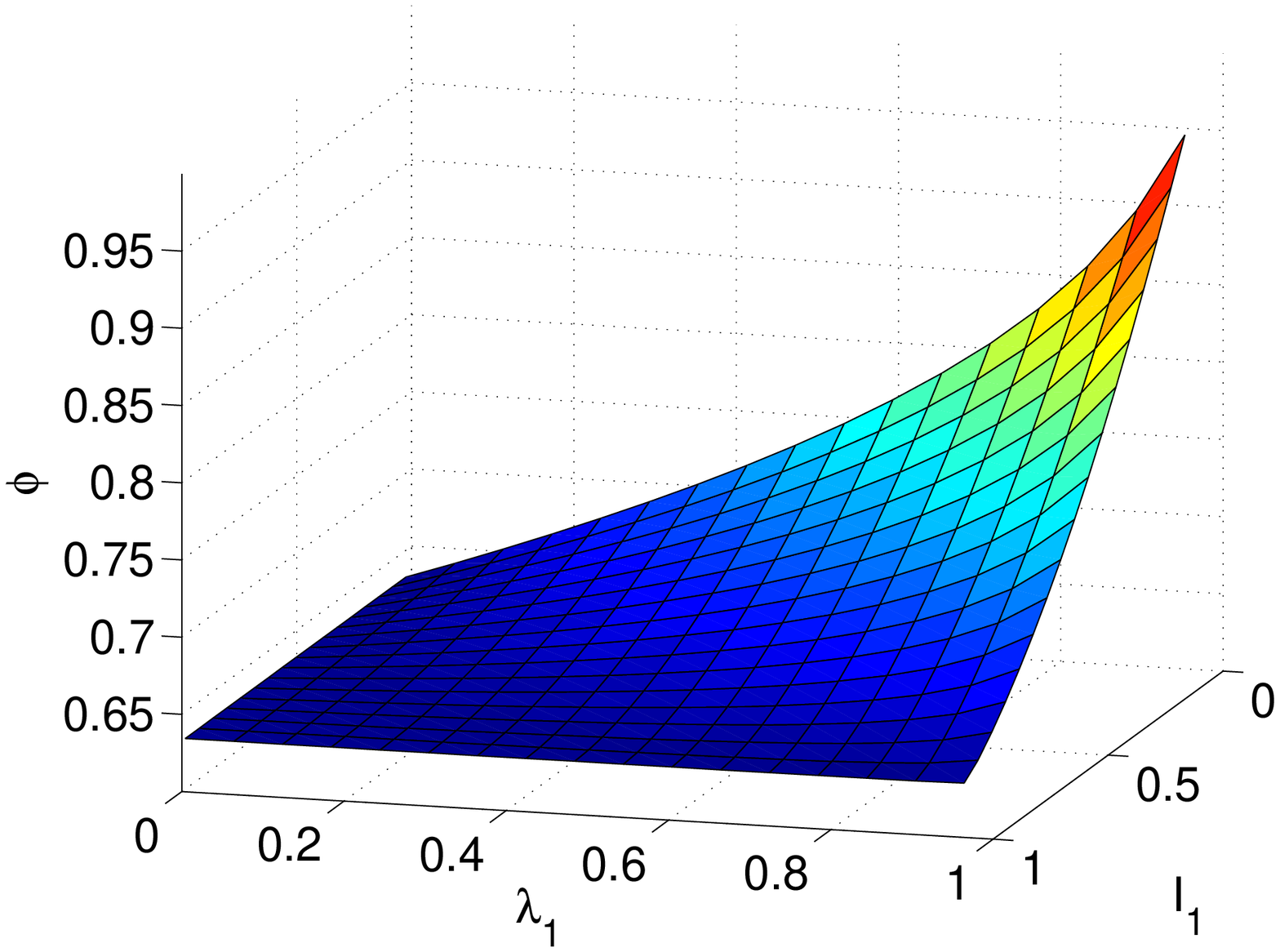}
\CAP{The average ratio of frozen agents $\phi(\lambda_1,I_1)$ calculated for the game of two groups ($ G=2 $) with $\alpha = 0.4$ and groups' impacts normalized as $ \MEANG{I_g^2} =1 $.}
\label{FIG.PHI}
\vfill
\includegraphics[scale=0.5,clip]{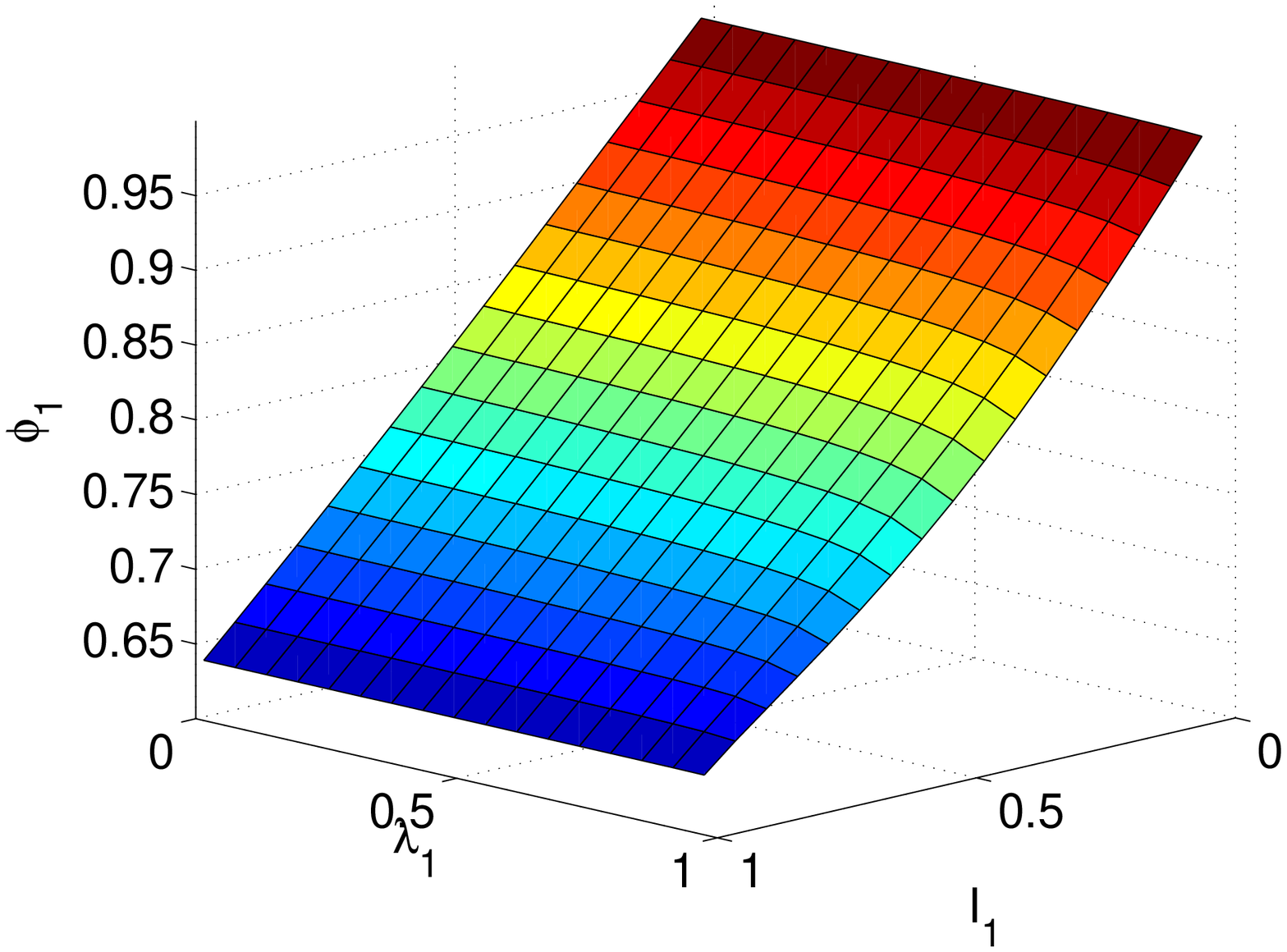}
\CAP{The ratio of frozen agents $\phi_1(\lambda_1,I_1)$ within the first group calculated for the game of two groups ($ G=2 $) with $\alpha = 0.4$ and groups' impacts normalized as $ \MEANG{I_g^2} =1 $.}
\label{FIG.PHI1} 
\end{figure}

\begin{figure}[h!t] \centering
\includegraphics[scale=0.5,clip]{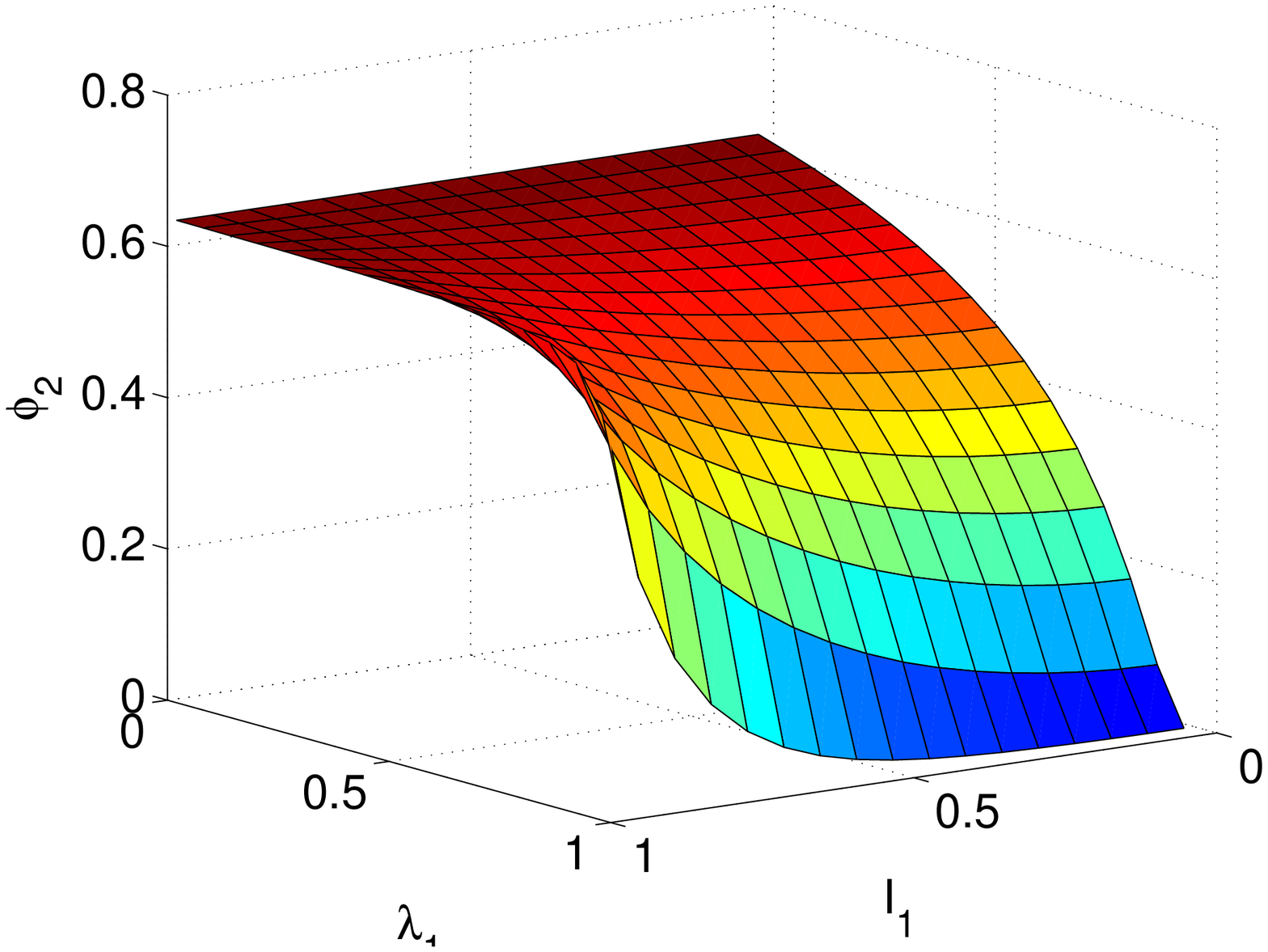}
\CAP{The ratio of frozen agents $\phi_2(\lambda_1,I_1)$ within the second group calculated for the game of two groups ($ G=2 $) with $\alpha = 0.4$ and groups' impacts normalized as $ \MEANG{I_g^2} =1 $.}
\label{FIG.PHI2}
\vfill
\includegraphics[scale=0.5,clip]{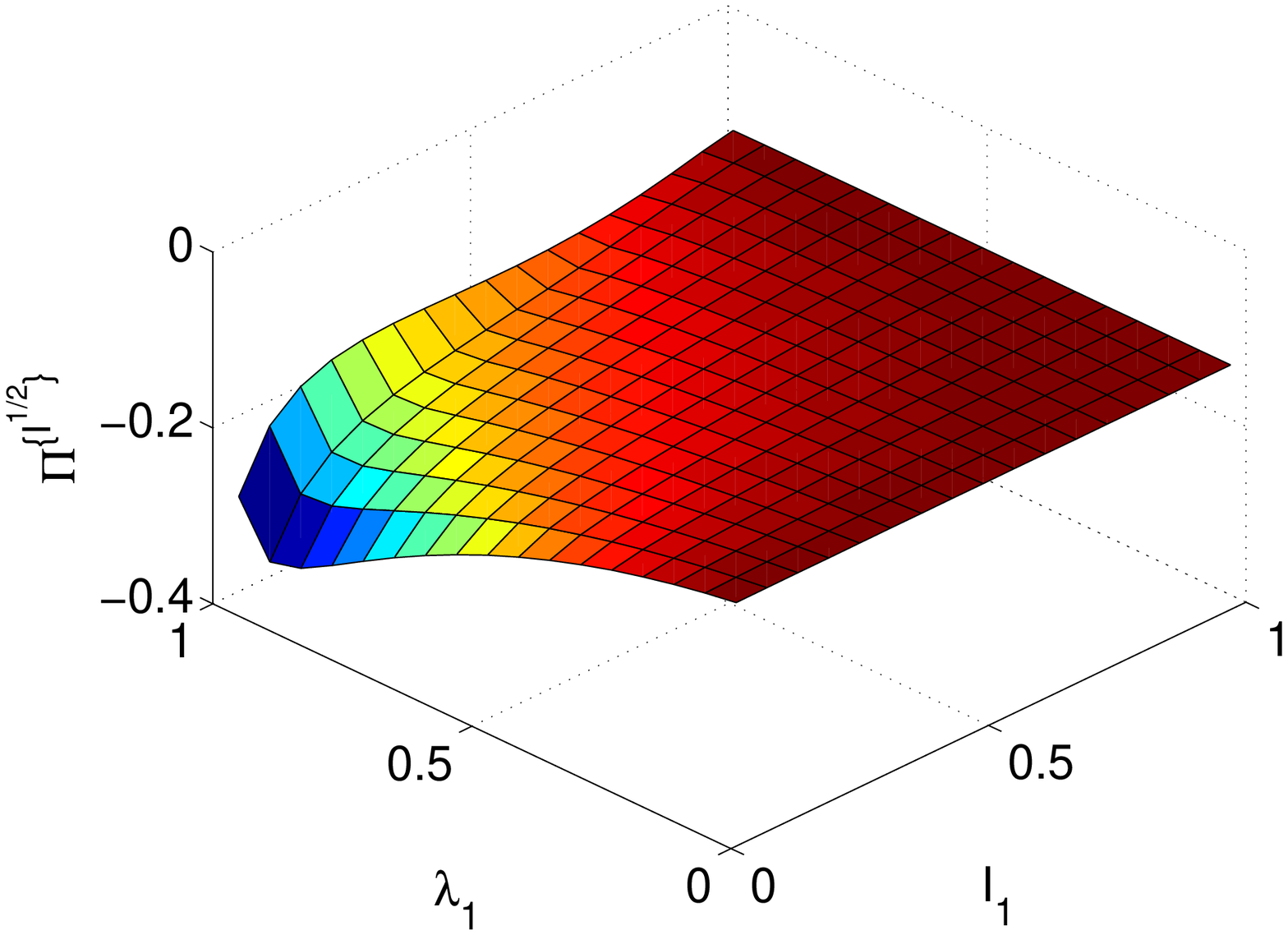}
\CAP{The average overall profit $\Pi^{\{I^{1/2}\}}(\lambda_1,I_1)$ calculated for the game of two groups ($ G=2 $) with $\alpha = 0.4$ and groups' impacts normalized as $ \MEANG{I_g^2} =1 $.}
\label{FIG.PI_1_2} 
\end{figure}

The next three figures depict the fraction of frozen agents. We plot their overall fraction $\phi$, fraction of frozen agents within the first group $\phi_1$ and fraction of frozen agents within the second group $\phi_2$, see Fig. \ref{FIG.PHI}, Fig. \ref{FIG.PHI1} and Fig. \ref{FIG.PHI2}, respectively. For higher ratio of frozen agents we observe also the higher overall outcome, compare $\phi$ with $\rho$. This was valid also in the canonical MG, where frozen agents were more effective on average. 

\begin{figure}[h!t] \centering
\includegraphics[scale=0.5,clip]{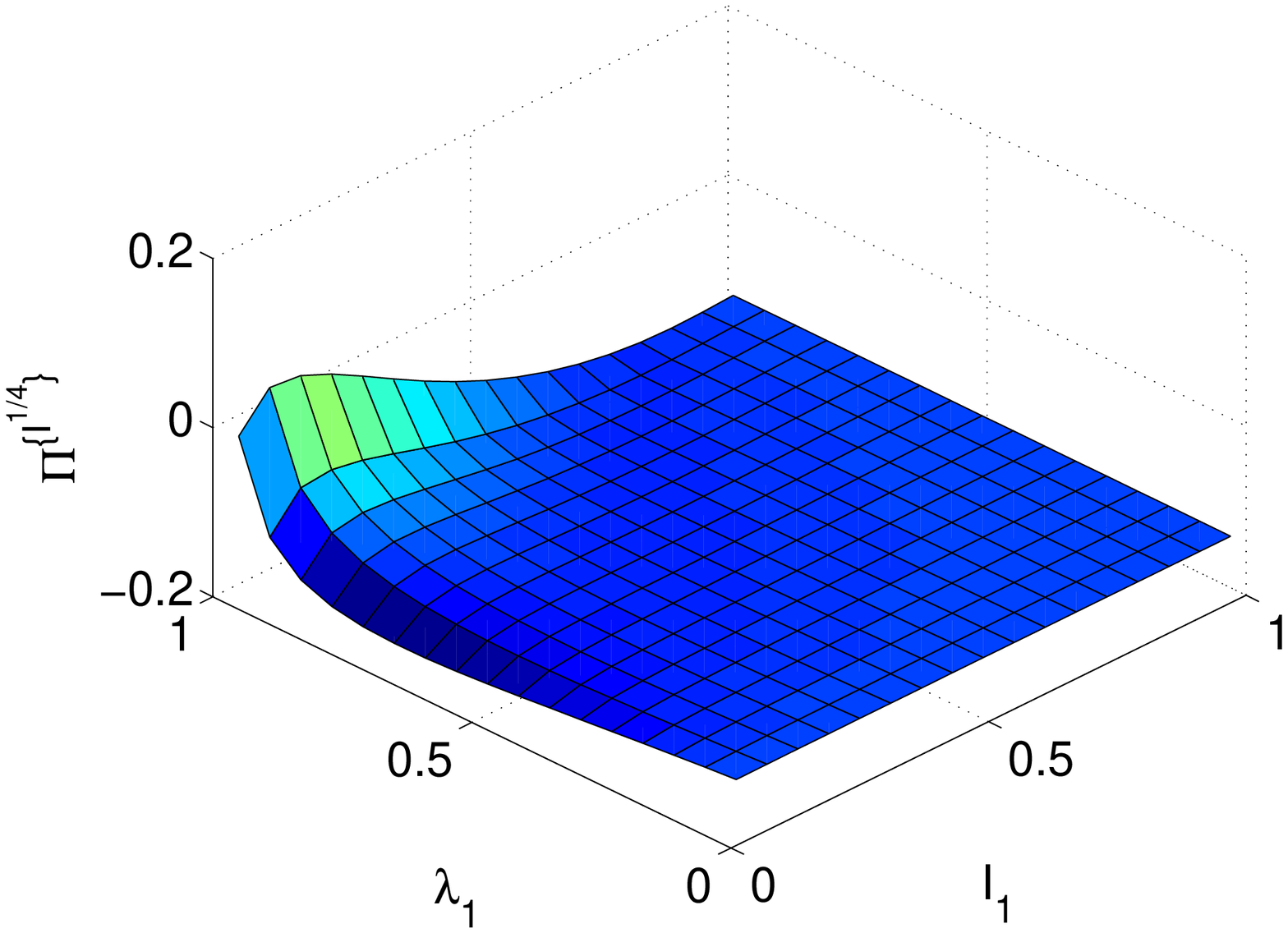}
\CAP{The average overall profit $\Pi^{\{I^{1/4}\}}(\lambda_1,I_1)$ calculated for the game of two groups ($ G=2 $) with $\alpha = 0.4$ and groups' impacts normalized as $ \MEANG{I_g^2} =1 $.}
\label{FIG.PI_1_4} 
\vfill
\includegraphics[scale=0.5,clip]{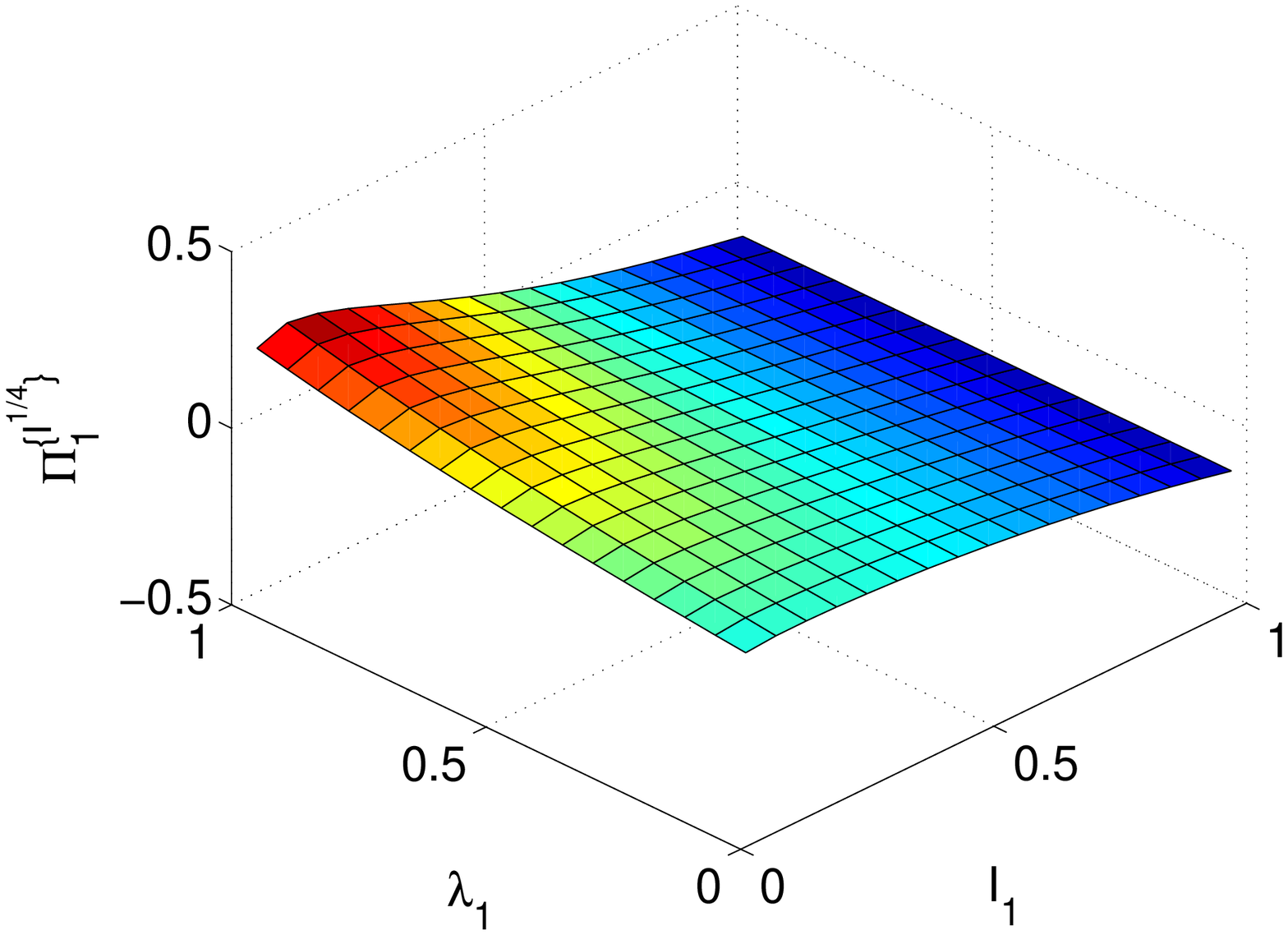}
\CAP{The average profit $\Pi_1^{\{I^{1/4}\}}(\lambda_1,I_1)$ of the first group calculated for the game of two groups ($ G=2 $) with $\alpha = 0.4$ and groups' impacts normalized as $ \MEANG{I_g^2} =1 $.}
\label{FIG.PI_1_4_1} 
\end{figure}

\begin{figure}[h!t] \centering
\includegraphics[scale=0.5,clip]{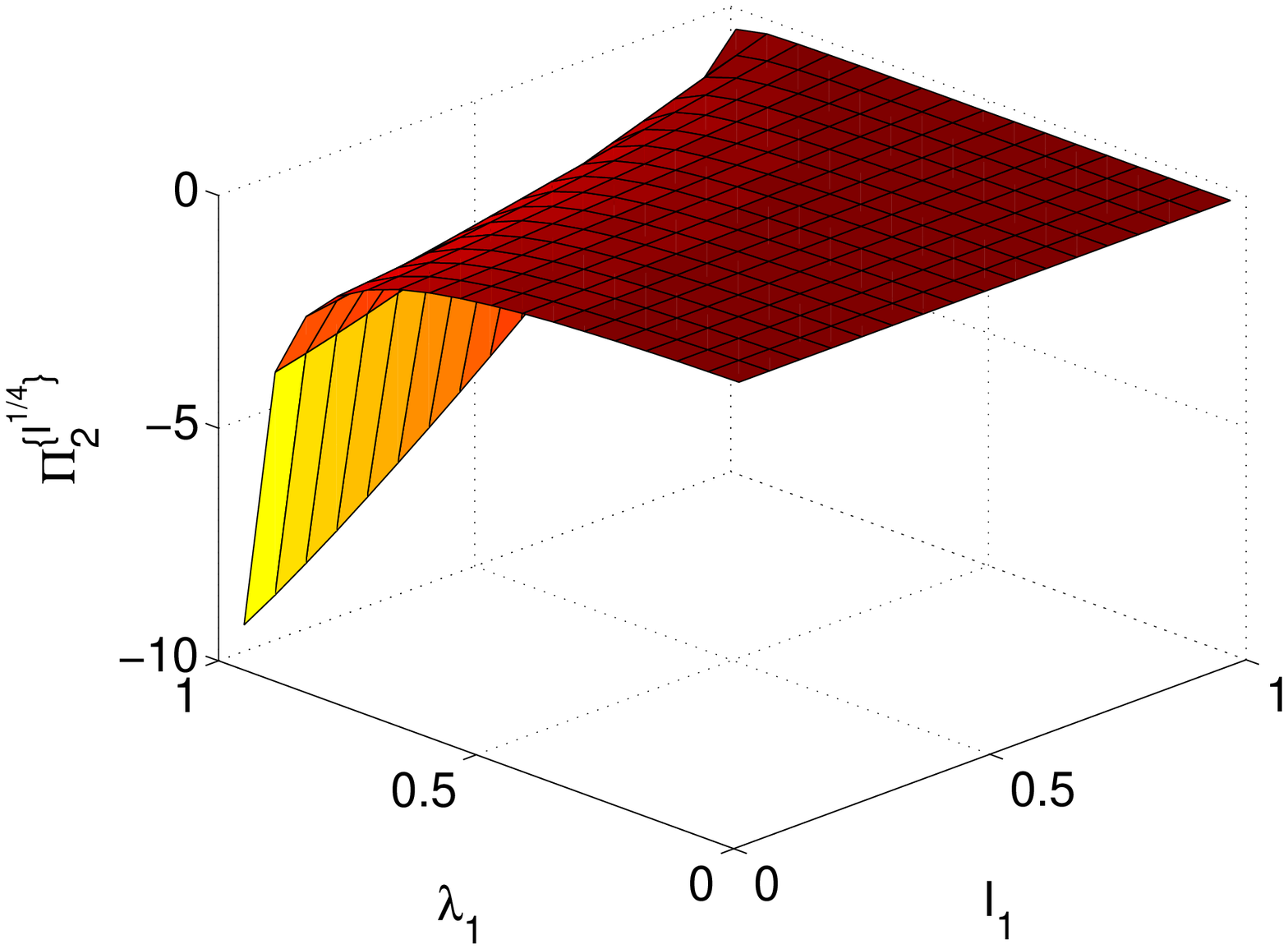}
\CAP{The average profit $\Pi_2^{\{I^{1/4}\}}(\lambda_1,I_1)$ of the second group calculated for the game of two groups ($ G=2 $) with $\alpha = 0.4$ and groups' impacts normalized as $ \MEANG{I_g^2} =1 $.}
\label{FIG.PI_1_4_2}
\vfill
\includegraphics[scale=0.5,clip]{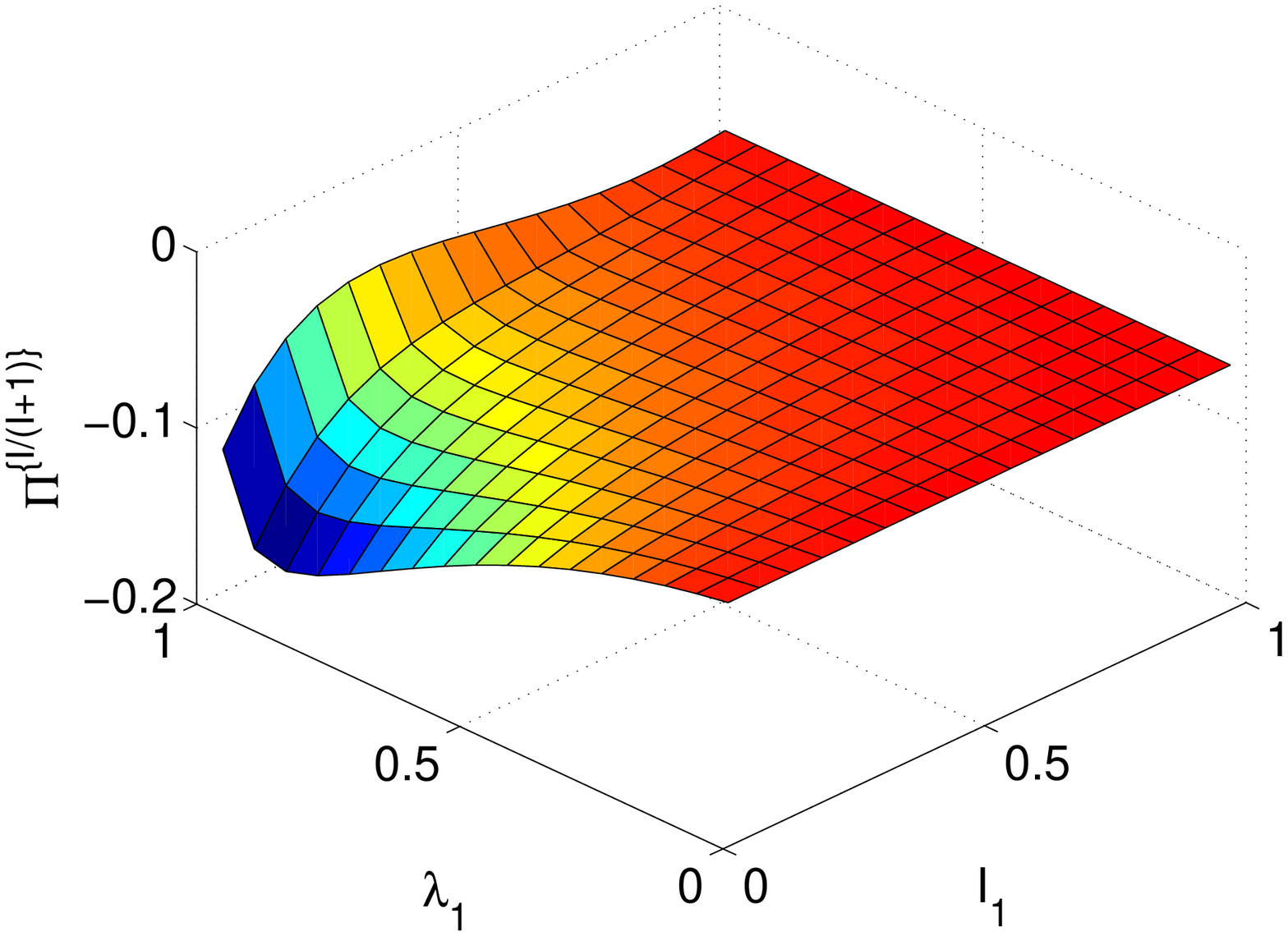}
\CAP{The average overall profit $\Pi^{\{I/(I+1)\}}(\lambda_1,I_1)$ calculated for the game of two groups ($ G=2 $) with $\alpha = 0.4$ and groups' impacts normalized as $ \MEANG{I_g^2} =1 $.}
\label{FIG.PI_2}
\end{figure}

In the last set of figures, we depict the way the overall average profit $ \Pi^{\{U(I)\}} $ 
is influenced by the particular
choice of the utility function $ U(I) $.  The overall average profits $\Pi^{\{I^{1/2}\}}$ and
$\Pi^{\{I^{1/4}\}}$ are 
depicted in Fig.  \ref{FIG.PI_1_2} and Fig.  \ref{FIG.PI_1_4}, respectively. For the utility 
function $U(I) = I^{1/4}$ we append Fig. \ref{FIG.PI_1_4_1} and
Fig. \ref{FIG.PI_1_4_2} depicting a separate average profits of both groups $\Pi_1^{\{I^{1/4}\}}$ 
and $\Pi_2^{\{I^{1/4}\}}$, respectively. Next, we add another overall average profit $\Pi^{\{I/(I+1)\}}$
using a different shape of the utility function, see Fig. \ref{FIG.PI_2}. For all these utility functions 
new local maxima of $ \Pi^{\{U(I)\}} $ appear. 

Although we did not solve explicitly the problem of maximizing the agents' utility, we consider the calculated dependence of the average profit on impacts as a guide for a hypothetical
choice of optimal impact. The location of the optimum depends crucially on the form of the utility function. Therefore, the introduction of the utility function is one of the cornerstones of 
our approach.

\section{Conclusion} 

We investigated in detail the modification of the Minority Game in
which the agents act on different scales. These heterogeneous scales
are implemented as agents' impacts and
can be interpreted in different ways. Within the traffic metaphor of the MG,
the impacts corresponds to the lengths of the vehicles the agents are driving. Alternatively, we can view it as a simplified method how to account 
for different frequencies at which the agents are playing. Then, an agent's impact is directly proportional to the inverse of the time scale at which the agent is participating. 

The agents are divided into several groups according to their
impacts. The MG model with heterogeneous impacts is solved by standard procedure
using the replica trick. The average outcome of each group was then
obtained, as well as the mutual predictabilities of the groups. To interpret 
properly the ensemble of outcomes for each group, we
introduce an utility function of the impact of agents. Within the
traffic metaphor, utility says how much profit can be expected having
vehicle of the given length. The average profit of individual groups 
is calculated from their average outcomes, with utility
considered.  We discuss the overall average profit of the system for the simplest 
case of two groups. The agents with lower impact are always in advantage, and the overall profit of the system can even be positive, which means that the 
MG viewed from this perspective may be a positive-sum game. This is in contrast with the
usual way of measuring the losses through the volatility which enforces that the MG is a 
negative-sum game.

Although we kept the traffic metaphor of the MG within this paper, we believe that the results  can be interpreted also in the language of stock markets. We argue that if we think of the impact of an agent as the size of the capital an investor is playing with, the relative outcome $\rho_g$ is more suitable description of an agent's profit. The real investors do compete for highest relative rather than absolute profit. We see that in absolute numbers the MG with heterogeneous impacts remains, of course, a negative-sum game, but subjectively felt relative profit of agents averaged over the whole system may be positive. This observation could give a cause for the heterogeneity of investors' sizes, which is one of the most fundamental features of the stock market \cite{gab_gop_ple_sta_03}. Interpreting the utility function $U(I)$ as an investor's subjective evaluation of the capital unit $ I $, the notion of $\Pi^{\{U(I)\}}$ is a tool suitable for finer analysis of this phenomena.

To sum up, by introduction diversity in the impact of agents, the
average profit in the MG, measured by an non-negative utility function,
may turn the MG into a positive-sum game, so that the diversity implies
advantage for all. We can conjecture that this is the driving
mechanism which favours emergence of groups acting with different
impacts. An important question which still remains to be solved is,
what would happen if each of the agents is free in her choice of the
impact. Both group sizes and group impacts would be free to be
adjusted to an optimum. The investigation of the properties of these
Nash equilibria is left to a future publication.

\section*{Acknowledgements} 

We want to thank Y.-C.Zhang for giving us the original motivation for this research.
We acknowledge constructive remarks of Ing. Miroslav K\'{a}rn\'{y}, DrSc., who helped us to improve
the manuscript. This work was carried our within the project AV0Z10100520 of the Academy 
of Sciences of the Czech republic and was  
supported by the M\v SMT of the Czech Republic, grant no. 
OC09078 and by  the GA \v{C}R grant no. 102/08/0567.

%
%

\end{document}